\documentclass[journal]{IEEEtran}

\ifCLASSINFOpdf

\else
 
\fi

\usepackage{amsmath}

\usepackage{graphicx}
\usepackage{bm}
\usepackage{cite}
\usepackage{caption}
\usepackage{subfigure}
\usepackage{array}
\usepackage{color}
\begin{document}

\title{REQA: Coarse-to-fine Assessment of Image \\ Quality to Alleviate the Range Effect}

\author{Bingheng~Li,
        and~Fushuo~Huo
        % and~Lihuo~He% <-this % stops a space
\thanks{B. Li is with the School of Electronic Engineering, Xidian University, Xian 710071, China and also with the Huawei Technologies Co., Ltd., Hangzhou 310000, China.
(email: bhlee@stu.xidian.edu.cn).}% <-this % stops a space
\thanks{F. Huo is with the Department of Computing, The Hong Kong Polytechnic University, Hong Kong, China (e-mail: hushuo.huo@connect.polyu.hk).}
}

\maketitle

\begin{abstract}
Blind image quality assessment (BIQA) of User Generated Content (UGC) suffers from the range effect, which indicates that on the overall quality range, mean opinion score (MOS) and predicted MOS (pMOS) are well correlated while focusing on a particular range, the correlation is lower. To tackle this problem,  a novel method is proposed from coarse-grained metric to fine-grained prediction. Concretely, we utilize global context features and local detailed features for the multi-scale distortion perception. Then, to further boost the ability of fine-grained assessment, we introduce the feedback mechanism, which is in accord with Human Vision System (HVS), to perceive detailed distortions gradually. Also, two coarse-to-fine loss functions are proposed to facilitate the feedback perception progress: a rank-and-gradient loss for coarse-grained metric keeps the assessment rank and gradient consistency between pMOS and MOS; a multi-level tolerance loss following the curriculum learning strategy is proposed to make a fine-grained prediction. Both coarse-grained and fine-grained experiments demonstrate that the proposed method outperforms the state-of-the-art ones, which validates that our method effectively alleviates the range effect. The codes are available at https://github.com/huofushuo/REQA.
\end{abstract}

\IEEEpeerreviewmaketitle

\section{Introduction}
\IEEEPARstart{I}{mage} quality assessment (IQA) explores how to imitate human beings to automatically assess image quality. Accurately describing the quality change has extensive applications in image restoration \cite{image_restoration}, image compression \cite{image_compression}, point cloud processing \cite{IEEETETCI3}, etc. IQA approaches can be generally divided into three categories: full-reference IQA approach (FR-IQA), reduced-reference approach (RR-IQA), and blind IQA approach (BIQA). FR-IQA and RR-IQA measure the similarity between the distorted image and reference image \cite{SSIM,VIF,FSIM,RR1,RR2,RR3}. However, in most authentic scenarios, it is hard to achieve ideal reference information. BIQA does not require any reference image as a prerequisite in predicting perceptual quality, so it has attracted great attention in recent years.

\begin{figure}[t]
\centering
\includegraphics[width=0.97\columnwidth]{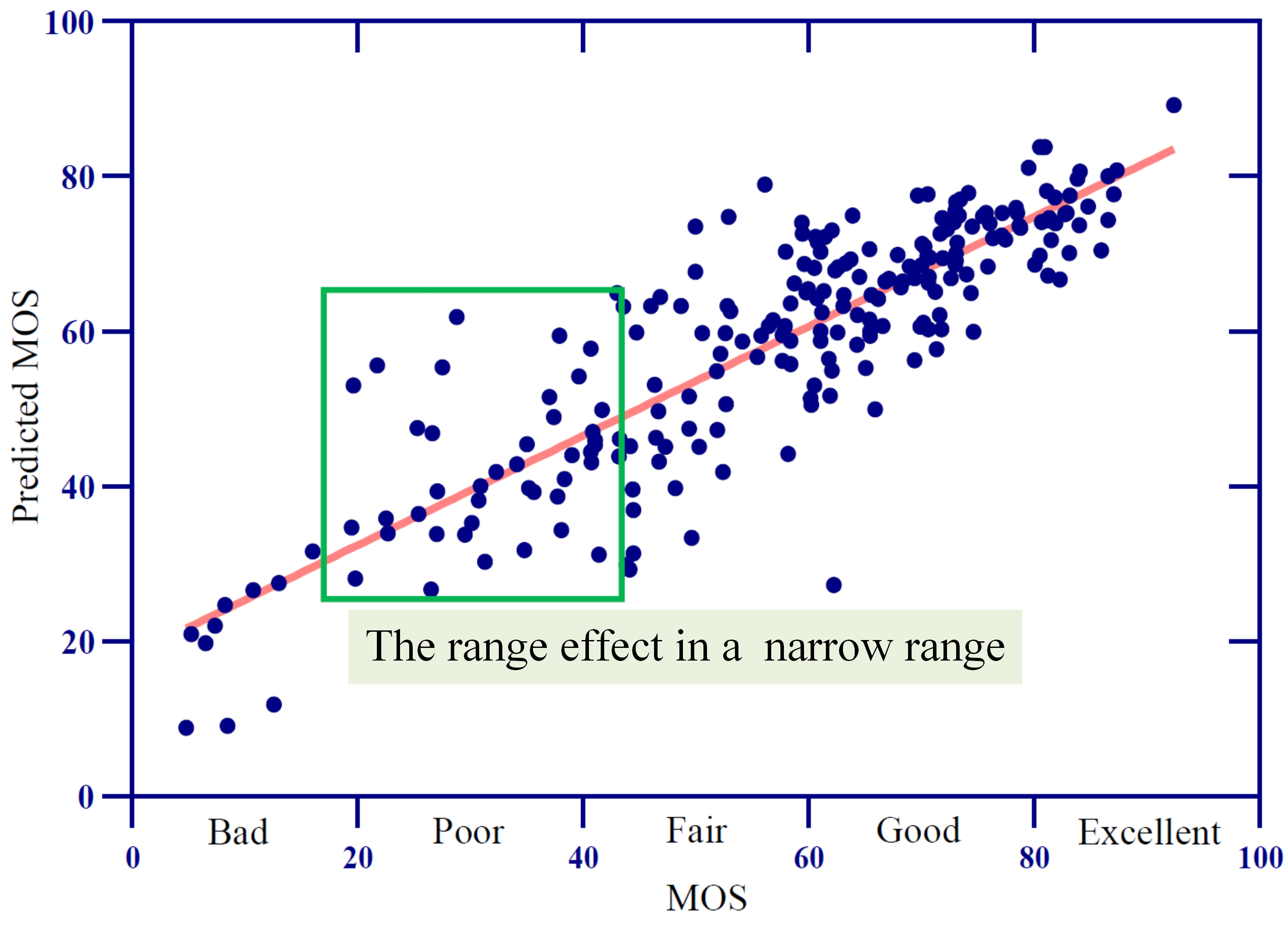}
\caption{Visualization of the predicted results of HyperIQA \cite{HyperIQA} on CLIVE\cite{LIVEC}. It depicts the \textbf{range effect}: In terms of the overall quality range, MOS and predicted MOS seem to be well correlated; While focusing on a particular range (e.g., points within the green box), the correlation is low.}
\label{fig20}
\end{figure}

% \begin{figure}[t]
% \centering
% \includegraphics[width=1\columnwidth]{1_1.pdf}
% \caption{Statistic results of the scale difference and the distance difference between pMOS and MOS on the authentic database.}
% \label{fig1}
% \end{figure}

Early BIQA methods\cite{CORNIA, DeepSim, WaDIQaM, MEON,  dipIQ,RRLRIQA, DeepQA,BPSQM, HVSIQA, HIQA, RAN4IQA, SIQA, RankIQA} mainly focus on synthetically distorted databases\cite{LIVEI,LIVEII,TID2013,CSIQ,KADID}, which consist of multiple distortions generated from limited scenarios. Since images captured in the real world suffer from ever-changing contents and more complicated distortions, accurately predicting their quality remains a challenge.
To deal with this problem, in recent years, some BIQA methods\cite{DBCNN, BLINDER, MetaIQA, SFA, IEEETETCI1, HyperIQA, NARCNN,  DeepRN, PQR, GraphIQA, IEEETETCI2, TMM} have been proposed, exploring novel learning strategies and leveraging complicated feature representation. However, although achieving relatively high correlations spanning a wide range from extremely bad to good on challenge datasets\cite{LIVEC,KonIQ, BID}, these approaches may be confronted with a drawback that they cannot perform well on a narrow quality range, which is referred as the range effect\cite{RE,range_effect1, REM}. Fig. 1 shows a visual example of the range effect of HyperIQA \cite{HyperIQA} on the CLIVE \cite{LIVEC} dataset.

How well BIQA methods perform on a narrow quality range is important for a realistic prediction of User Generated Content (UGC). According to \cite{PaQ, range_effect1}, the distributions of image quality in the authentically distorted databases are narrow and peak as compared to those synthetic ones, as most pictures captured in the real world are improved by the imaging device. According to the distribution property, the streaming media servers can screen the images with extreme qualities to obtain images with moderate qualities. These images will be integrated as the benchmark of subsequent image enhancement and image-effect synthesis. Thus, we claim that the BIQA model should possess more a fine-grained perception ability to assess image quality, alleviating the range effect.

Some methods\cite{REM, range_effect1} proposed a new evaluation criterion aiming at FR-IQA, which eliminates the range effect to some extent. However, there are few attempts to solve this problem in BIQA yet. Recently, Zhang et al.\cite{FGIQA} revisited the IQA and conducted a survey on the fine-grained IQA. They pointed out that existing IQA methods do not address the potential fine-grained IQA. In this paper, we first attempt to develop a more fine-grained blind image \underline{Q}uality \underline{A}ssessment method to alleviate the \underline{R}ange \underline{E}ffect, termed as REQA. Our method follows the coarse-to-fine principle. 
For the \textbf{coarse-grained} assessment, 
existing metric-learning-based methods \cite{DeepSim}\cite{PieAPP}\cite{RankIQA} map the samples into the multi-dimensional feature vectors, as for clustering or nearest neighbor classification utilizing a distance function that captures pair-wised similarity\cite{ML}. This is effective on synthetic images. These
methods map the synthetically distorted images into the feature vectors which represent distortion classification or distortion level. However, in authentic databases, images possess nonidentical scenes and immeasurable distortions that increase the difficulty to obtain the coherence of feature distributions\cite{KonIQ}. Even if it can be achieved between a pair of images. it may not be quality-aware. The proposed method maps a set of images to supervised scores, based on these scores, a rank-and-gradient metric can be conducted as for reducing the prediction deviation from a wide quality rank and further alleviate the order confusion of the predicted quality sequence. 
For the \textbf{fine-grained} assessment, which is unsolved for existing methods, we adopt a multi-stage prediction strategy. Previous BIQA approaches adopt a one-time strategy to predict MOS with a feedforward structure. However, the way neglects the feedback mechanism of the Human Visual System (HVS) as it is important for perceptual learning \cite{PL}. Neurologically speaking, feedforward hierarchy underlies implicit processing for initial vision at a glance, and feedback connections add details to explicit vision with scrutiny\cite{Feedback1}. The same applies to BIQA where fine-grained cognition of image quality is achieved through feedback processing where high-level and low-level features are recurrently integrated by HVS. Moreover, the feedback-based learning approach has been proven more effective than the commonly employed feedforward paradigm in prediction tasks\cite{FeedBack2}. In this paper, MOS prediction is constantly refined under multi-level tolerance constraints through a feedback structure. Besides, the coarse-grained metric is fused into the structure as the prior knowledge as it is easier compared to fine-grained prediction from the perspective of curriculum learning\cite{CL}.

In summary, our contributions are four-fold:
\begin{itemize}
\item To our best knowledge, we take the first attempt to develop a fine-grained blind image quality assessment method to alleviate the range effect.
\item We propose the effective coarse-to-fine strategy, which not only utilizes global context features and local detailed features for the multi-scale distortion perception but also introduces the feedback mechanism to perceive detailed distortions gradually.
\item We also devise two coarse-to-fine loss functions to facilitate the feedback perception progress.
\item Comprehensive experiments based on traditional coarse-grained evaluation and fine-grained evaluation show the effectiveness of our method.
\end{itemize}

\section{Related Work}
In this section, as we try to handle the range effect in BIQA, we give the detailed review of BIQA methods for synthetically distorted images and authentically distorted images, respectively.
\subsection{BIQA for Synthetically Distorted Images}
Previous research of BIQA approaches for the synthetic task mainly follows two kinds of ideas: traditional methods and learning-based methods.

Commonly traditional BIQA approaches first extract hand-crafted features based on the empirical analysis and then adopt a regression function to map the features into the quality score. The most well-known category is based on the characteristics of natural statistical scenes (NSS), such as DIIVINE\cite{DIIVINE}, BLIINDS-II\cite{BLIINDSII} and BRISQUE\cite{BRISQUE}. This type of method assumes that natural images have certain statistical properties affected by  distortion, which would make the image look unnatural. Therefore, features can be extracted from frequency, spatial, and wavelet domains based on the statistical properties of an image to predict its quality score. In addition, there are also some other methods based on the human visual system (HVS), such as NRSL\cite{NRSL} and RISE\cite{RISE}. These methods utilize HVS to construct the quality-aware features, assuming that HVS is adapted to the structure information. However, these hand-crafted features are time-consuming and meantime lack of generalization ability due to the diversity of image contents and distortions.

Unlike the traditional BIQA methods, the learning-based BIQA approaches automatically generate quality-aware features. In the early stage, CBIQ\cite{CBIQ} and CORNIA\cite{CORNIA} introduced the code-book feature-based learning into BIQA. These methods first utilize raw image patches extracted from a set of unlabeled images to learn a dictionary in an unsupervised manner, and then encode the test images on the dictionary to obtain the feature representations for quality estimation. In the subsequent development, CNN pushes the significant development of BIQA thanks to its powerful learning ability. 
% Kang et al.\cite{Kang} propose a shallow CNN network which performs the mapping progress from a predicted image to its perceptual quality. 
WaDIQaM\cite{WaDIQaM} proposes a significantly deeper framework that comprises ten convolutional layers and five pooling layers for feature extraction, and two fully connected layers for regression. As these networks grow deeper and wider, they need larger annotated databases for training. Due to extremely labor-intensive and costly subjective experiment, the current IQA databases are too small to meet this requirement.

To deal with the small sample problem, it needs to explore more effective learning strategies and leverage more complicated features. Some methods find a way out in transfer learning. MEON\cite{MEON} and DBCNN\cite{DBCNN} pretrain a classification model on a large-scale synthetic database to acquire the initialized network parameters which are, to some extent, distortion-aware. RankIQA\cite{RankIQA} and dipIQ\cite{dipIQ} propose a pairwise learning-to-rank (L2R) algorithm, Which can learn to rank images in terms of image distortion. Then, they transfer the prior knowledge learned from ranked images to a traditional CNN. After fine-tuning it on IQA databases, it can improve the accuracy of IQA. RRLRIQA\cite{RRLRIQA} makes further exploration and models the BIQA as a Markov decision process to optimize the whole image-quality ordering directly. During training phase, not all distortions
or images are handled equally well, \cite{TMM} improves recent methods with the online hard example mining strategy.

Since humans are the ultimate receivers of images, the properties of the human visual system (HVS) should also be modeled in a data-driven manner. Perceptual error map is learned to guide quality prediction in \cite{DeepQA}\cite{BPSQM}, where DeepQA\cite{DeepQA} is designed from FR-IQA methods, and BPSQM\cite{BPSQM} utilizes the U-Net to generate a similar map of the distorted image for reference. In HVS-Net\cite{HVSIQA}, visual saliency and just noticeable difference (JND)\cite{JND} are taken into account to acquire the perceptually important features. Meantime, the rank loss is proposed to penalize the model when the order of its predicted quality scores is biased against that of the ground truth scores.

Some GAN-based methods have also been developed in the last few years \cite{GAN1,HIQA,RAN4IQA,AIGQA, ICME}. H-IQA\cite{HIQA} and RAN4IQA\cite{RAN4IQA} suppose that HVS unveils the mask of distortion and recreates a hallucinated scene without distortion in mind. In addition, AIGQA\cite{AIGQA} proposes an active inference module based on the generative adversarial network (GAN) to predict the primary content. Since these IQA methods have achieved great improvement in synthetically distorted databases, a drawback exists when applied to authentic ones. The reference information is inevitably utilized in their training stage, which makes them limited in user generated content (UGC) due to the lack of reference images.

\subsection{BIQA for Authentically Distorted Images}
Most BIQA methods focus on synthetically distorted images, but relatively few approaches have been proposed to deal with the more challenging problem of authentic IQA. In recent years, based on multiple learning strategies, some methods are proposed to cope with this challenge. BLINDER\cite{BLINDER} and DBCNN\cite{DBCNN} pretrain a classification model on  photographically generated classification databases such as ImageNet to acquire the quality-aware network parameters, which can help the regression task in IQA databases. MetaIQA\cite{MetaIQA} adopts model-agnostic meta-learning (MAML)\cite{MAML} to learn the prior knowledge among different synthetic distortions. However, due to the imbalance between the synthetic distortion and the authentic distortion, the learned knowledge cannot be generated effectively to the authentically distorted IQA databases, which is obvious in the experiment results. SFA-IQA\cite{SFA} and HyperIQA\cite{HyperIQA} make IQA models understand the content diversity in authentic databases. The former utilizes semantic feature aggregation (SFA) to eliminate the impact of image content variation, and the latter utilizes a hyper-network architecture to evaluate the image quality adaptively according to the image content. In NAR-CNN\cite{NARCNN}, the authors propose a dual-path network to support IQA from a reference image with a similar scene but is not aligned. Considering the fact that an image receives divergent subjective scores from different human raters, PQR\cite{PQR} and DeepRN\cite{DeepRN} utilize the distribution of subjective scores to describe image quality. GraphIQA \cite{GraphIQA} develop Distortion Graph Representation (DGR) learning framework for BIQA, in which each distortion is represented as a graph and GraphIQA distinguishes distortion types by learning the contrast relationship between these different DGRs

In this paper, as for the unanswered range effect in BIQA of UGC, we propose the novel REQA method, which has a more fine-grained ability to alleviate the range effect. Compared to the state-of-the-art methods, REQA can effectively tackle the prediction deviation in narrow ranges on the authentic databases, which contain images close to user generated content.

\begin{figure}[t]
\centering
\includegraphics[width=1\columnwidth]{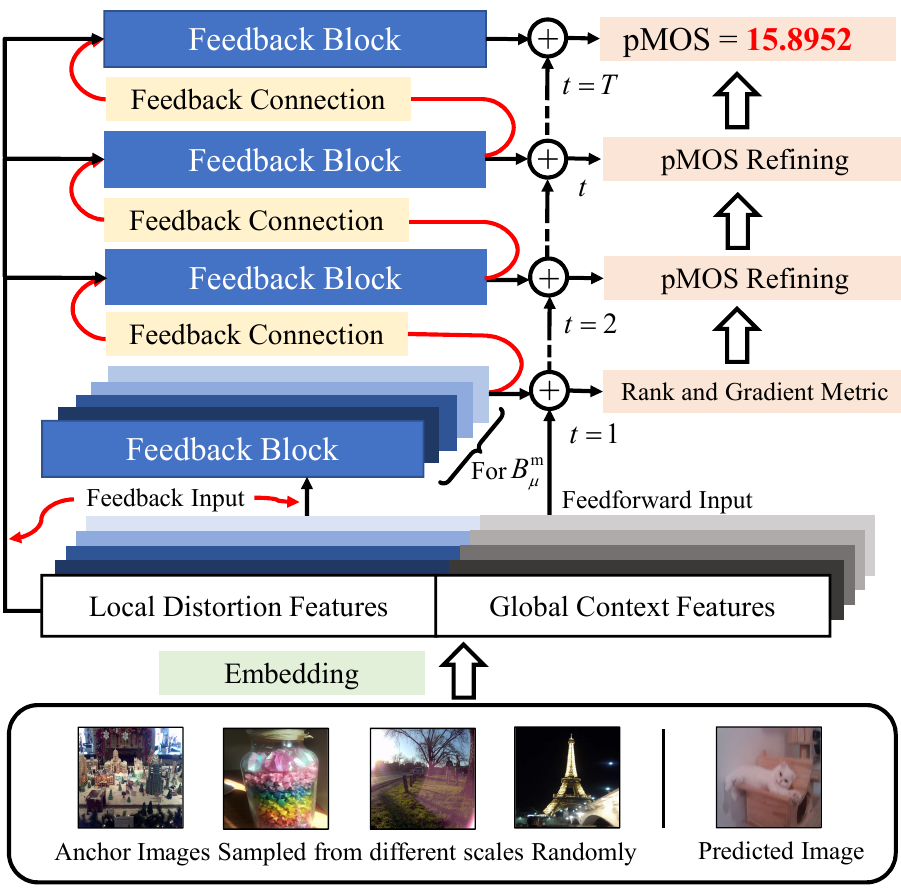} % Reduce the figure size so that it is slightly narrower than the column. Don't use precise values for figure width.This setup will avoid overfull boxes.
\caption{Flowchart of the proposed REQA. REQA adopts a feedback hierarchy to realize coarse-to-fine quality assessment within multiple time steps. Here, a rank-and-gradient metric accomplishes coarse-grained assessment, and multi-stage pMOS refinements complete fine-grained assessment. In addition, REQA integrates image feature representations from two aspects, where the first is to process multi-scale distortion features through iteration and the second is the fusion of context features from Transfomer Encoder\cite{TE}.}
\label{fig2}
\end{figure}

\section{Proposed Method}
To alleviate the range effect and improve the prediction performance in a narrow quality range, a novel BIQA method (named REQA) is proposed. As is illustrated in Fig. 2, REQA is end-to-end trainable and divides and conquers the BIQA task to multi-time steps with the feedback structure, realizing the coarse-to-fine image quality assessment. As for more quality-aware feature representations, we add a light-weight Transformer Encoder\cite{TE} on the top level to obtain the non-local features. In this section, we first discuss the network framework from two aspects: the backbone feedback network for multi-scale distortion perception and the Transformer Encoder for context understanding. Then we introduce two coarse-to-fine loss functions.

\subsection{Feedback Network for Multi-scale Distortion Perception}
\begin{figure*}[t]
\centering
\includegraphics[width=1\textwidth]{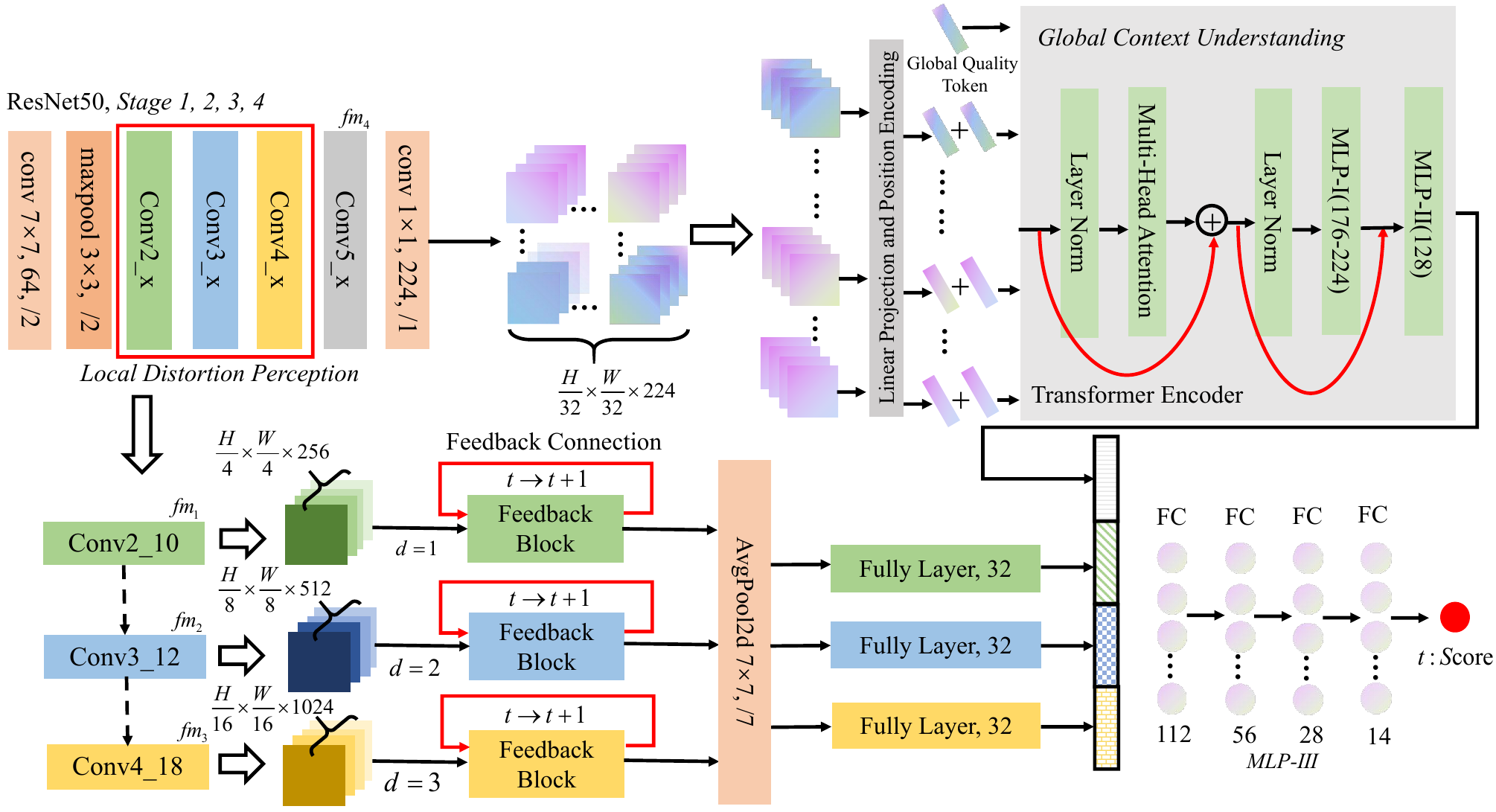} % Reduce the figure size so that it is slightly narrower than the column.
\caption{The overview framework of REQA. It consists of three parts as the baseline network:ResNet-50}\cite{ResNet}, feedback network (FN), and Transformer Encoder (TE). Here, as the core part of REQA, FN processes multi-scale features iteratively utilizing feedback blocks. TE provides the non-local representation for the outputs of feedback blocks in each time step. In addition, global and local representations are firstly concatenated and then mapped by multilayer perceptron(MLP-\uppercase\expandafter{\romannumeral3}) into a supervised score to predict image quality.
\label{fig1}
\end{figure*}
Feedback Network (FN) is the core part of the proposed method, which controls the feedback process to perceive image degradation from multi-scale distortion features continually. The whole network consists of three elements, including feature extraction, time-domain iteration, and space-domain integration. The detail of this module is shown in Fig. 3.

\subsubsection{Feature extraction} Multi-scale features contain diverse low-level information, which has been proved to be effective for IQA\cite{DeepSim}\cite{HyperIQA}.  Following \cite{HyperIQA, SGD, GraphIQA}, REQA adopts the ResNet-50\cite{ResNet} as the backbone to acquire multi-scale features. Specifically, we remove global average pooling layers and fully connected layers of ResNet-50, and initialize corresponding network parameters using pretrained model in ImageNet. Finally, multi-scale features $fm_{1}, fm_{2}$, and $fm_{3}$ are extracted from $conv2\_{10}$, $conv3\_{12}$, $conv4\_{18}$ layers, respectively.

\subsubsection{Time-domain iteration} Meantime, neurology \cite{Feedback1} also proves that HVS can add details into distortion areas through feedback connection\cite{FeedBack2}. The property of HVS contributes to quality perception. Thus, multi-scale features combined with a feedback mechanism improve the fine-grained ability of BIQA. Local distortion perception is completed by three feedback blocks. Each block takes a single scale feature map, and the feedback feature extracted from the last time step is fed to the current step as input. Through integrating and correlating these features, quality-aware representation is got to finish the assessment task of the current step. For feedback block $d$, this iterative process is expressed as:

\begin{equation}
\begin{array}{l}
fo_{d,t} = \mathcal{G}_{d}(fm_{d}, fh_{d, t-1}; \phi_{d}) \\[3mm]
d=1,2,3, t = 1, 2, \cdots, T
\end{array}
\end{equation}
where $fh_{d, t-1}$ is hidden state in time step $t-1$, $fo_{d,t}$ is the output in time step $t$, $\phi_{d}$ represents the block parameters, and $\mathcal{G}_{d}$ expresses the mapping function.

\subsubsection{Space-domain integration} Space-domain integration is dependent on a feedback block whose inner structure is based on ConvLSTM\cite{ConvLSTM}, as illustrated in Fig. 4. An LSTM cell uses multiple gates to control information saving, discarding , merging, and finally uses hidden states to pass feedback through iterations. We briefly present the connections between gates in the LSTM cell as followings:
\begin{equation}
\begin{array}{l}
fi_{d, t}=\sigma(W_{d,x_{i}}fm_{d}+W_{d,h_{i}}fh_{d,t-1}) \\[3mm]
fg_{d,t}=\sigma(W_{d,x_{f}}fm_{d}+W_{d,h_{f}}fh_{d,t-1}) \\[3mm]
fo_{d,t}=\sigma(W_{d,x_{o}}fm_{d}+W_{d,h_{o}}fh_{d,t-1}) \\ [3mm]
\tilde{fc}_{d,t}=tanh(W_{d,x_{c}}fm_{d}+W_{d,h_{c}}fh_{d,t-1}) \\ [3mm]
fc_{d,t}=f_{t}^{d}\circ fc_{d,t-1}+fi_{d,t}\circ \tilde{fc}_{d,t}\\[3mm]
fh_{d,t}=o_{t}^{d}\circ tanh(fc_{d,t})
\end{array}
\end{equation}
where $fi_{d,t}$ is the input gate, $fg_{d,t}$ is the forget gate, $fo_{d,t}$ is the output gate, $fc_{d,t}$ is the memory cell,  $\sigma$ is the logistic sigmoid function, and $W_{*}$ is the weight matrix of conv block. In the proposed method, according to $d$, we set three conv blocks whose details are also shown in Fig. 4. The output $fo_{d,t}$ is finally mapped into $\bm{\upsilon}_{d,t}\in R^{1\times 32}$.

\begin{figure*}[t]
\centering
\includegraphics[width=16cm]{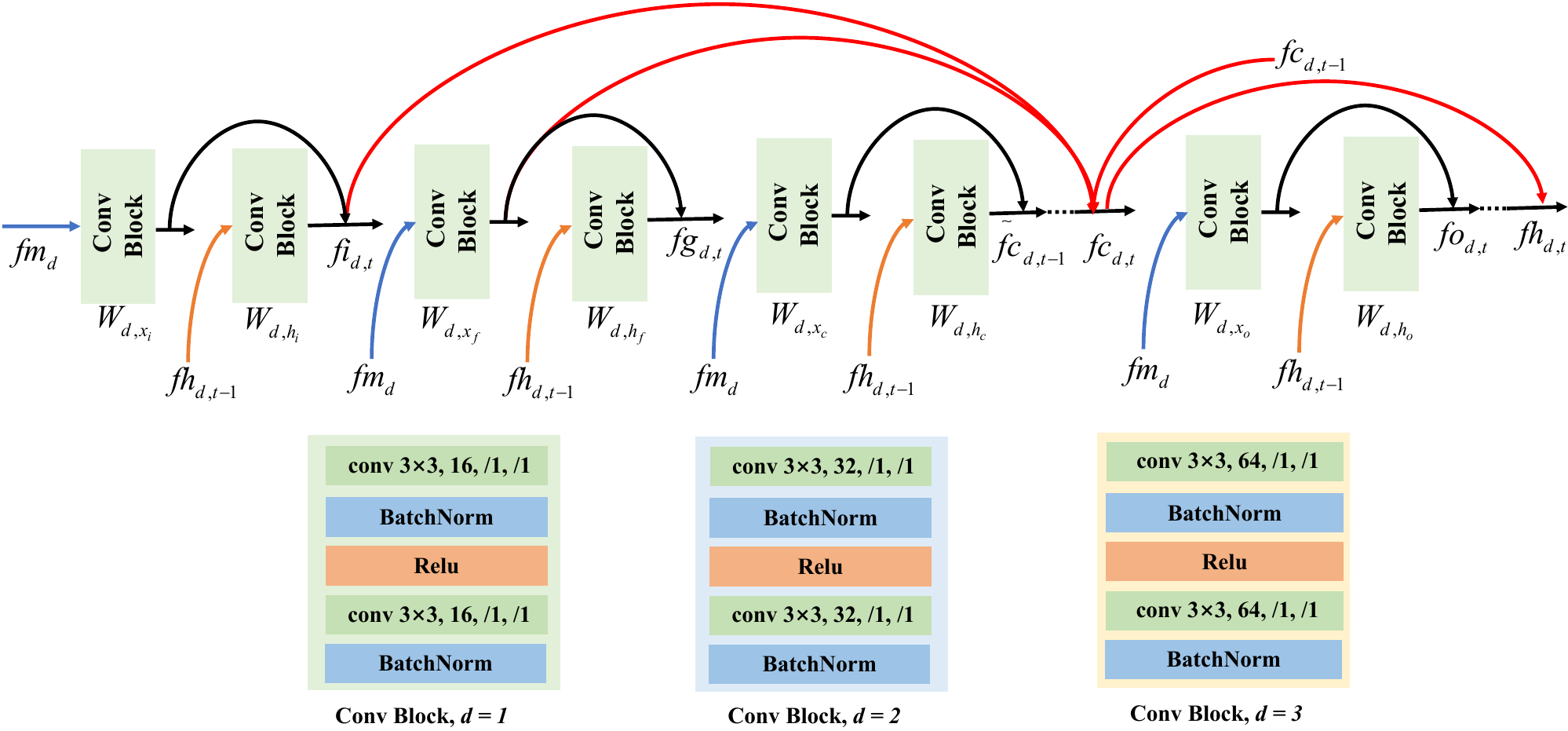}
\caption{Architecture details of feedback block.}
\label{fig4}
\end{figure*}

\subsection{Transformer Encoder for Context Understanding}

To some extent, the perception of the local distortion can assess the degree of image quality degradation effectively. However, there are still some situations where local distortion cannot be quality-aware. For example, photographers are used to \textcolor{black}{throwing} the background out of focus to improve \textcolor{black}{the} visual effect of the foreground. In this case, when we only pay attention to the background and ignore its correlation with the foreground, it is easily regarded as \textcolor{black}{a} fuzzy distortion that \textcolor{black}{affects} the image quality. Thus, adding the context information to local feedback features contributes to understanding image quality comprehensively.

Compared to traditional CNN, Transformer\cite{TE} manages to capture \textcolor{black}{long-range} interactions thanks to its multi-head attention mechanism, that is, it is effective to obtain non-local features of an image. Meanwhile, CNN has some inherent inductive biases, such as translation \textcolor{black}{invariant}, scale \textcolor{black}{invariant,} and so on, which \textcolor{black}{are} not possessed by Transformer. These properties make CNN suitable for feature extraction. However, the layer occupies more computation resources compared to CNN. Therefore, we only apply \emph{one} transformer layer on the top level of ResNet-50 to acquire the feedforward context feature. The details of the Transformer Encoder (TE) \textcolor{black}{are} illustrated in Fig. 3. It consists of three elements: patch embedding, multi-head attention module\textcolor{black}{,} and multilayer perceptron.

\subsubsection{Patch embedding} The original distorted image $I_{i}\in {R}^{H\times W\times 3}$ is processed by \textcolor{black}{ResNet-50,} and respective feature map $fm_4\in {R}^{\frac{H}{32}\times\frac{W}{32}\times C}$ is acquired by $conv5\_9$. Here, $(H, W)$ is the resolution of $I_{i}$\textcolor{black}{,} and $C$ is the number of channels.
Transformer encoder takes \textcolor{black}{a} one-dimensional vector as input. As for this, $fm_4$ is cut into $N$ patches firstly\textcolor{black}{,} where each patch $\mathcal{P}_{n,i}\in {R}^{P\times P \times C}$, $n\in[1,N]$ and $N=H\cdot W/ (32^2 \cdot P^2)$. Then, $\mathcal{P}_{n, i}$ is flattened into a vector $\bm{\eta}_{n,i}\in {R}^{1\times P^2 C}$\textcolor{black}{,} which is mapped into $\bm{\zeta}_{n,i}\in {R}^{1\times D}$ by a learnable matrix $W_{pe}$ as:
\begin{equation}
\bm{\zeta}_{n,i} = \bm{\eta}_{n,i} \cdot W_{pe}, W_{pe}\in R^{P^2 C\times D}
\end{equation}
where, in the experiment, we set $D=224$. Moreover, to keep positional information, a standard learnable position embedding $\bm{\rho}_{n, i}\in {R}^{1\times D}$ is added into $
\bm{\zeta}_{n,i} $ to obtain the embedding $\bm{\varepsilon}_{n,i}$ of $\mathcal{P}_{n,i}$ as:
\begin{equation}
\bm{\varepsilon}_{n,i}=(\bm{\zeta}_{n,i}  + \bm{\rho}_{n, i})^T
\end{equation}

To realize the perception of the whole quality representation, we concatenate a trainable token $\bm{x}_{quality} \in {R}^{1\times D}$ to the patch embeddings $\{\bm{\varepsilon}_{1,i}, \bm{\varepsilon}_{2,i}, \cdots , \bm{\varepsilon}_{N,i}\}$, similar to BERT's\cite{BERT} and ViT's\cite{VIT} class token. Finally, the input matrix ${X}_{i}$ is defined as:
\begin{equation}
{X}_{i}=[\bm{x}_{quality}^{T}; \bm{\varepsilon}_{1,i}; \bm{\varepsilon}_{2,i}; \cdots; \bm{\varepsilon}_{N,i}]^{T}
\end{equation}
\subsubsection{Multi-head attention} \textcolor{black}{Following \cite{TE},} self-attention map is acquired through a weighted sum over all values $V$ of the input matrix ${X}_{i}$. Each element in the weighting matrix $A$ are the pairwise correlation between two patch representations in ${X}_{i}$\cite{TE}\textcolor{black}{,} which is calculated by the dot product of\textcolor{black}{ the} respective query and key. The specific computational process is as:

\begin{equation}
\begin{array}{lr}
[Q,K,V]=\mathcal{L}(X_{i})W_{sa}, W_{sa}\in R^{D \times 3D_{h}} \\
A=softmax(\frac{QK^{T}}{\sqrt{D_{h}}}), A\in R^{(N+1) \times (N+1)} \\
\mathcal{S}(X_{i})=AV
\end{array}
\end{equation}
where  $\mathcal{L}(\cdot)$ denotes layer normalization and $\mathcal{S}(\cdot)$ is self-attention operation.

Multi-head attention utilizes multiple self-attention operations to integrate global context relevance from local patch embeddings. Here, we set $D/D_h$ self attention as:

\begin{equation}
\mathcal{M}(X_{i})=\mathcal{L}([\mathcal{S}(X_{1});\mathcal{S}(X_{2}); \cdots; \mathcal{S}(X_{D /D_{h}})]W_{ma} + X_{i})
\end{equation}
where $W_{ma}\in R^{D \times D}$ is the transition matrix.

\subsubsection{Multilayer perceptron}  We extract a 1D vector $\xi_{quality}$ from $\mathcal{M}(X_{i})$ which \textcolor{black}{corresponds} to $\bm{x}_{quality}$. \textcolor{black}{Then} MLP-\uppercase\expandafter{\romannumeral1} takes $\xi_{quality}$ as the input to obtain $\tilde{\xi}_{quality}$. After MLP-\uppercase\expandafter{\romannumeral2} processing $\tilde{\xi}_{quality}$, $\bm{\upsilon}_{m}$ is computed. Finally, through feature concatenating and mapping, $\mathcal{F}_{t}(I_{i};\theta)$ is achieved as:
\begin{equation}
\mathcal{F}_{t}(I_{i};\theta)=MLP-\uppercase\expandafter{\romannumeral3}(\bm{\upsilon}_{m}\oplus \bm{\upsilon}_{1,t}\oplus \bm{\upsilon}_{2,t}\oplus \bm{\upsilon}_{3,t})
\end{equation}
where $\oplus$ represents \textcolor{black}{concatenation} operation.

\subsection{\textcolor{black}{Coarse-to-Fine Loss Functions}}
The proposed method employs a multi-time strategy $t \in \{1, 2,..., T\}$ where \textcolor{black}{the} feedback mechanism controls the progress of \textcolor{black}{the} BIQA task. In particular, it captures multi-scale quality-aware features, and then they are integrated and mapped into the quality score that is finally utilized by each time step to accomplish the assessment subtask. The whole assessment process is coarse-to-fine, containing the coarse-grained metric and fine-grained prediction.
 
\subsubsection{Coarse-grained metric} As for coarse-grained metric, the proposed method redesigns the \textcolor{black}{sampling} strategy of training data. We randomly sample a mini-batch with $K$ 2-tuples $B=\{(I_{k},s_{k})\}_{k=1}^{K}$ from \textcolor{black}{the} current training set. Here, $I_{k}$ and $s_{k}$ are the training images and MOS labels, respectively. This randomly sampled mini-batch contains training samples from multiple quality scales. Coarse-grained metric aims to improve the perception of quality difference, so we group the training samples to form a micro-batch $B_{\mu}^{m} = \{(I_{i}, s_{i})\}_{i=1}^{5}$ where a predicted image and four anchors are sampled from five different quality scales. Here, $m\in[1,M]$ where $M = \lfloor K/5 \rfloor$. Compared to the state-of-the-art ones\cite{DBCNN}\cite{SFA}\cite{HyperIQA}\cite{PQR}, the \textcolor{black}{sampling} strategy has two advantages as it can be adaptive to the metric task, and moreover, it can fix the small sampling problem in BIQA because of the combinatorial diversity.

The realization of coarse-grained metric is based on \textcolor{black}{quality ranking} and gradient keeping, which occupies the first-time step of feedback learning as the prior knowledge of fine-grained prediction. \textcolor{black}{Different from \cite{HVSIQA, tcsvt5}, the proposed method puts forward a loss which can not only metric quality order but also keep the distance difference of \textcolor{black}{the} pairwise predicted scores consistent with one of respective ground truth scores.} For $I_{i}$ and $I_{j}$, rank loss is first defined as:

\begin{equation}
\footnotesize
\begin{aligned}
L_{t=1}^{rank}(i,j) = max(0, \frac{-(s_{i}-s_{j})\times(\mathcal{F}_{t=1}(I_{i};\theta)-\mathcal{F}_{t=1}(I_{j};\theta))}{\vert |s_{i} - s_{j} |\vert_1 + \sigma}) 
\\ \times(\vert |s_{i} - s_{j} |\vert_1 + 1)
\end{aligned}
\end{equation}
where $\theta$ is the network parameters, $\mathcal{F}_{t=1}(I_{i};\theta)$ represents the quality prediction of image $i$ in $t=1$  and $\sigma$ is a small stability term. In our experiment, we set $\sigma=0.0001$. When the order of pairwise prediction sequence is consistent with the ground truth order, rank loss \textcolor{black}{equal} zero, otherwise\textcolor{black}{,} it is reduced to an absolute loss as:
\begin{equation}
L_{t=1}^{rank}(i,j) = \vert |s_{i}-s_{j}| \vert_1 \times \vert |\mathcal{F}_{t=1}(I_{i};\theta)-\mathcal{F}_{t=1}(I_{j};\theta)| \vert_1
\end{equation}

The rank loss keeps the order consistency, and gradient loss maintains the stability of pair-wised quality distance difference as:
\begin{equation}
L_{t=1}^{gradient}(i,j) = \Big| \vert |s_{i} - s_{j}|\vert_{1} -  \vert |\mathcal{F}_{t=1}(I_{i};\theta)-\mathcal{F}_{t=1}(I_{j};\theta) |\vert_{1} \Big |
\end{equation}

Overall, for a micro batch $B_{\mu}^{m}$, coarse-grained loss is defined as:
\begin{equation}
L_{t=1}^{Coarse}=\sum_{i=1}^{5}\sum_{j=i+1}^{5}(L_{t=1}^{rank}(i,j) + L_{t=1}^{gradient}(i,j))
\end{equation}

\subsubsection{Fine-grained prediction} Fine-grained prediction is inspired by the episodic curriculum learning\cite{CL}, adopting an easy-to-hard strategy to realize $t \in[2,T]$ prediction refinement. Specifically, We set a different threshold for each time step to compute the loss, and as the feedback information is processed continually, the threshold is reduced. Overall, the fine-grained loss is defined as:
\begin{equation}
\begin{array}{lr}
L_{t}^{Fine}=max(0, \vert |s_{i} - \mathcal{F}_{t_{i}}(I_{i};\theta)|\vert_1 - l_{t_{i}}) \\[5mm]
l_{t} > l_{t+1}\geq 0, t < t+1
\end{array}
\end{equation}
where $L_{t}^{Fine}$ is the loss function in time step $t$, and $l_{t}$ represents the threshold in $t$. The whole loss of the proposed method is calculated as:
\begin{equation}
\begin{array}{lr}
L = w_{1}L_{t=1}^{Coarse} + w_{2}L_{t=2}^{Fine} + w_{3}L_{t=3}^{Fine} + \cdots + w_{T}L_{t=T}^{Fine}\\[3mm]
w_{1}+w_{2}+\cdots+w_{t}=1
\end{array}
\end{equation}

In our experiment, we set $T=4$, $l_{2} = 5$ , $l_{3} = 2.5$  and $l_{4} = 0$. The values of $w_{1}, w_{2}, w_{3}$ and $w_{4}$ is dynamically adjusted, \textcolor{black}{with the multi-time training task going on,} to satisfy the curriculum learning principle. The set for $w_{1}, w_{2}, w_{3}$ and $w_{4}$ is in Section \uppercase\expandafter{\romannumeral4}-A(3).

\section{Experiment}
In this section, we first introduce the experimental protocols, including databases, criterion\textcolor{black}{,} and implementation details. Then we compare REQA with the state-of-the-art BIQA methods \textcolor{black}{in terms of fine-grained prediction as well as traditional coarse-grained prediction} performance\textcolor{black}{,} respectively. \textcolor{black}{Next, we implement} a series of ablation experiments to verify the contribution of different components of REQA. Finally, we also present some visualization samples acquired from the fine-grained distortion perception module to verify the effectiveness of the feedback hierarchy.

\subsection{Experiment Protocols}
\subsubsection{Databases}
Three authentically distorted IQA databases\textcolor{black}{,} including CLIVE\cite{LIVEC}, KonIQ-10k\cite{KonIQ}, and BID\cite{BID} are used to evaluate the performance of \textcolor{black}{the} proposed method. CLIVE contains 1162 images captured from  \textcolor{black}{diverse} mobile devices under \textcolor{black}{real-world} conditions. KonIQ consists of 10073 images, selected from 10 million YFCC100M entries. The sampling strategy of it concerns the authenticity of distortions, the diversity of content, and quality-related indicators. BID comprises 586 images with realistic blur distortions such as motion blur and defocusing blur, etc.

These databases are constructed based on quality \textcolor{black}{ratings}, i.e., bad, poor, fair, good, and excellent. Crowdsourcing strategy utilizes subjects to give multiple quality ratings to \textcolor{black}{a} single image. These ratings \textcolor{black}{are} then rescaled into $[1,100]$ to compensate for the biases of individual evaluations. Here, the higher score corresponds to higher quality. \textcolor{black}{Through} averaging them, MOS/DMOS is obtained as the quality label.
In our experiment, because these three databases provide no original category labels, we divide the ground truth data according to Absolute Category Rating (ACR)\cite{BT500} into five equal potions as Excellent $(80-100)$, Good $(60-80)$, Fair $(40-60)$, Poor $(20-40)$ and Bad $(0-20)$ as for \textcolor{black}{fine-grained} prediction performance experiment.

\subsubsection{Criterion}
\textcolor{black}{To evaluate the fine-grained ability to alleviate the range effect as well as the traditional coarse-grained ability,} the Speraman Rank Order Correlation Coeffcient (SROCC) and the Pearson Linear Correlation Coefficient (PLCC) \textcolor{black}{are utilized as the evaluation metrics.} SROCC is to measure the \textcolor{black}{monotonicity} between the ground truth data and the prediction scores. PLCC is to evaluate the linear correlation between these two. Given $N$ images, the SROCC is defined as:
\begin{equation}
	SROCC = 1 - \frac{6\sum_{i=1}^{N} d_{i}^{2}}{N(N^2-1)}
\end{equation}
\textcolor{black}{where $d_i$} is the rank difference between MOS and pMOS of the $i-th$ image. And PLCC is computed as:
\begin{equation}
	PLCC=\frac{\sum_{i=1}^{N} (s_{i}-\mu_{s_{i}})(\tilde{s}_{i}-\mu_{\tilde{s}_{i}})}{\sqrt{\sum_{i=1}^{N} (s_{i}-\mu_{s_{i}})^2 }\sqrt{\sum_{i=1}^{N} (\tilde{s}_{i}-\mu_{\tilde{s}_{i}})^2}}
\end{equation}
where $s_{i}$ and $\tilde{s}_{i}$ \textcolor{black}{denote} MOS and pMOS of the $i-th$ image, and $\mu_{s_{i}}$, $\mu_{\tilde{s}_{i}}$ correspond to the mean of each.
\subsubsection{Implement Details}
We conducted \textcolor{black}{both} training and testing using Pytorch on \textcolor{black}{an} NVIDIA 2080 Ti GPU. In the proposed model, \textcolor{black}{all the training images are deployed traditional data augment strategy, i.e. randomly crop the images to $224 \times 224$ pixel patches like \cite{NARCNN, DBCNN, MetaIQA, HyperIQA}.} The results are obtained from 20 train-test iterations. In each iteration, we randomly select $80\%$  images for training, and the remaining $20\%$ for testing, so there is no overlap between the training set and the test set. We train our model using Adam optimizer with weight decay $5e-4$ for 40 epochs. Learning rates for the backbone \textcolor{black}{ResNet-50} and the other modules are first set to $2e-5$ and $2e-4$\textcolor{black}{,} respectively, and reduced by 10 in $10th$ epoch, $20th$ epoch, and $30th$ epoch\textcolor{black}{,} respectively.

In addition, the setting for $w_{1}$, $w_{2}$, $w_{3}$\textcolor{black}{,} and $w_{4}$ is consistent with the episodic curriculum learning\cite{CL}. First of all, we prioritize the assessment tasks corresponding to $w_{1}$, $w_{2}$, $w_{3}$\textcolor{black}{,} and $w_{4}$ in Eq.6 based on the easy-to-hard strategy. Specifically, the coarse-grained metric occupies the highest priority; \textcolor{black}{Meantime}, fine-grained loss with a bigger threshold possesses higher priority. \textcolor{black}{Besides}, we divide the training into four stages, with ten \textcolor{black}{epochs} for each stage. Finally, the setting is $w_{1}=0.25$, $w_{2}=0.5$, $w_{3}=0.25/3$, $w_{4}=0.25/3$ in stage 1, $w_{2}=0.5$, $w_{3}=0.25$, $w_{4}=0.25$ in stage 2, $w_{2}=0.25$, $w_{3}=0.5$, $w_{4}=0.25$ in stage 3 and $w_{1}=0.25$, $w_{2}=0.25$, $w_{3}=0.25$ and $w_{4}=0.5$ in stage 4.

\begin{figure*}[t]
    \centering
        \subfigure[MetaIQA\cite{MetaIQA}]{\includegraphics[width=0.4\linewidth]{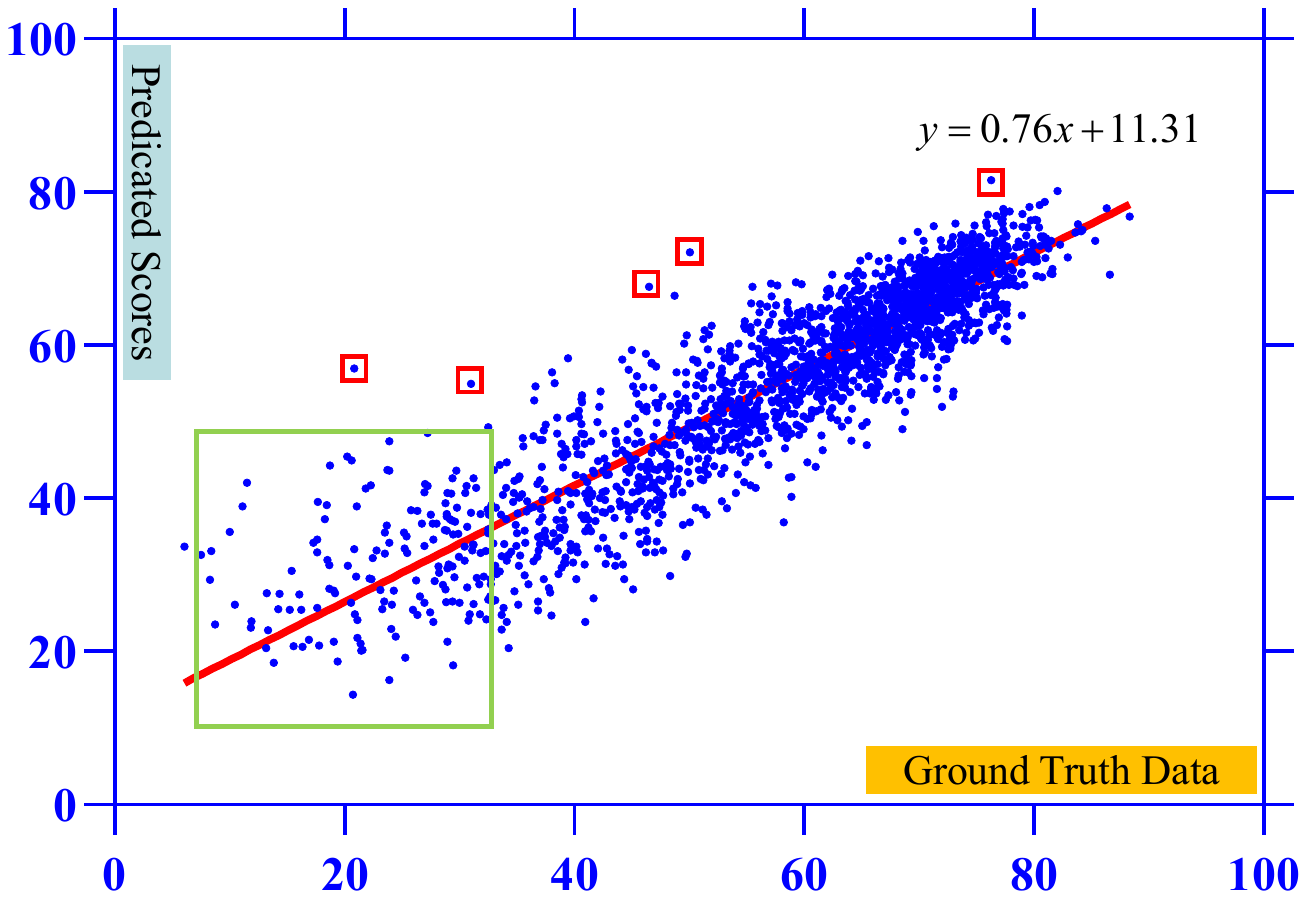}}
        \subfigure[HyperIQA\cite{HyperIQA}]{\includegraphics[width=0.4\linewidth]{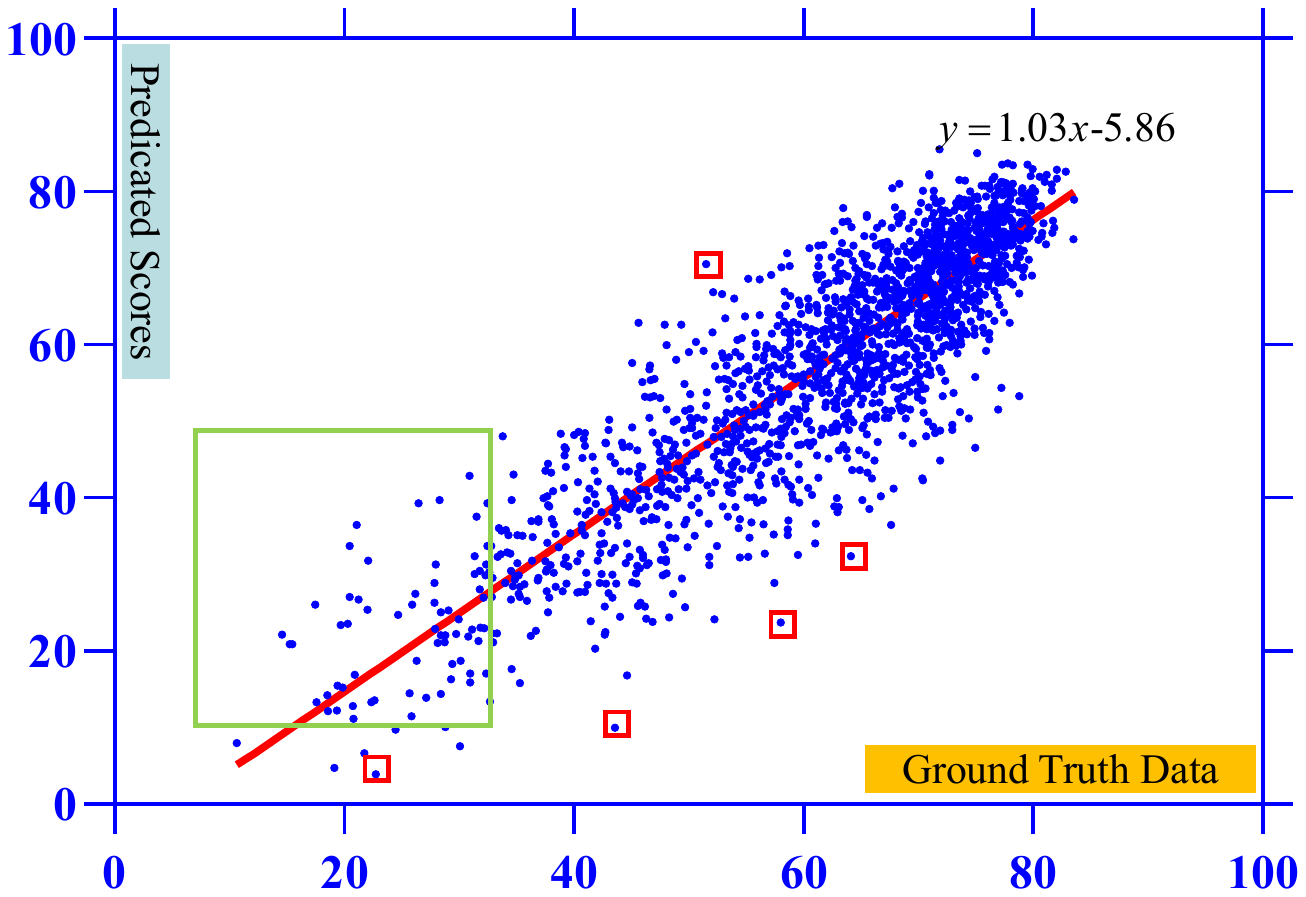}} \\
        \subfigure[GraphIQA\cite{GraphIQA}]{\includegraphics[width=0.4\linewidth]{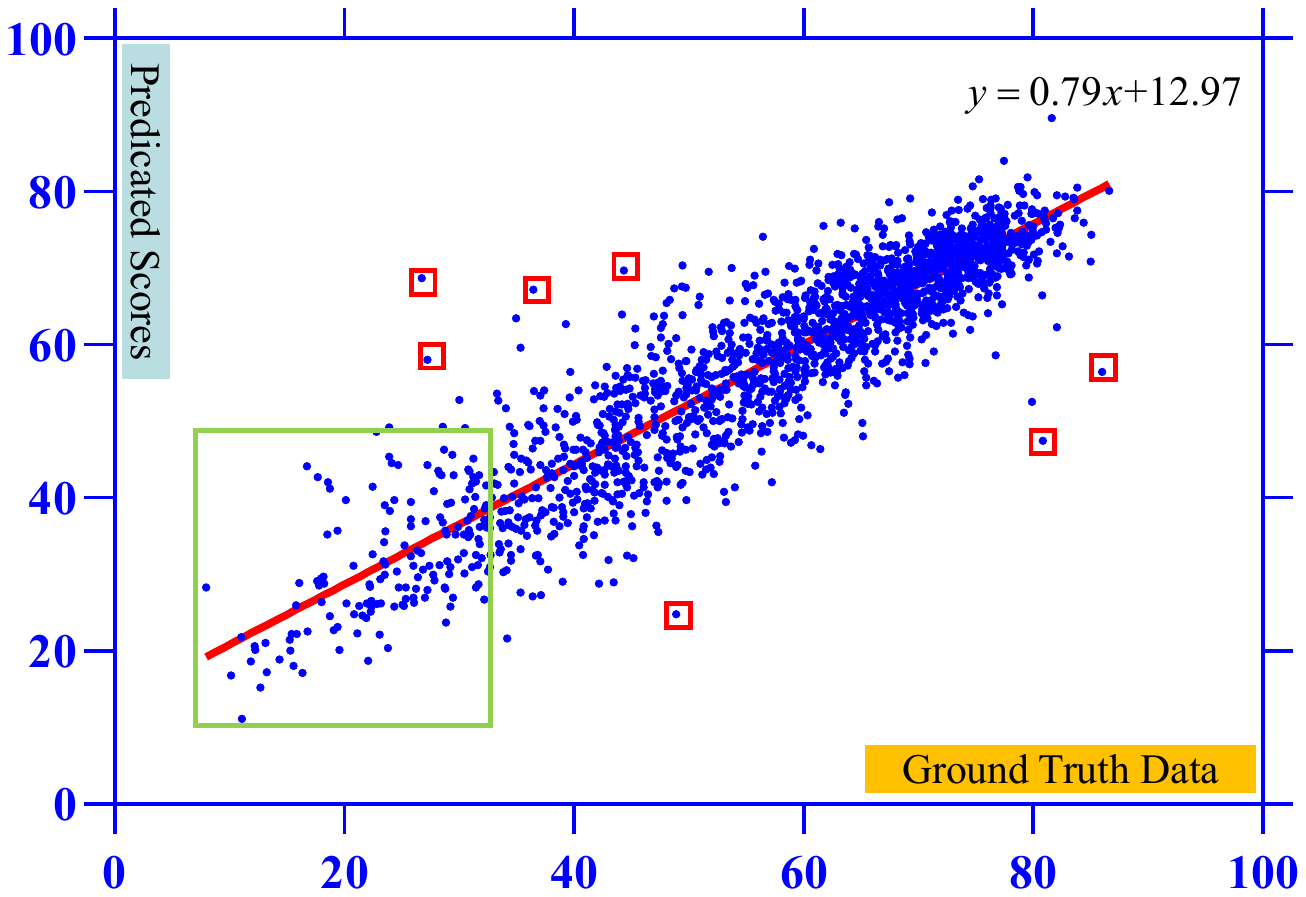}}
        \subfigure[REQA]
        {\includegraphics[width=0.4\linewidth]{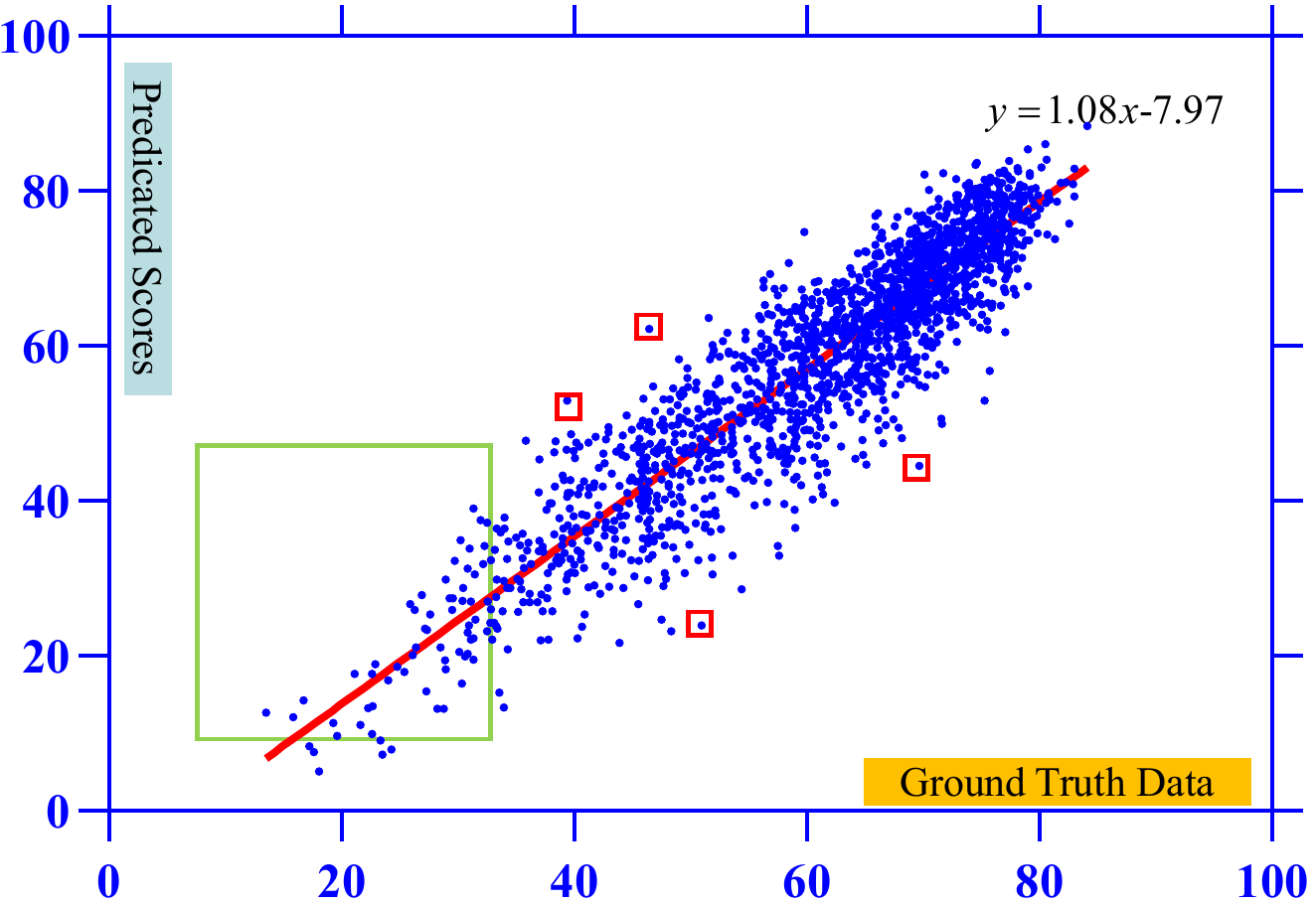}}\\

        \caption{Visualization of predicted results. \textcolor{black}{Green boxes focuses on the particular range and red boxes concerns the outliers} .}%The experimental results for $R_t= \exp \left( \left[1, -0.5, -2\right]^T \right)$.
\end{figure*}

\subsection{\textcolor{black}{Fine-grained } Prediction Performance Experiment}
In this \textcolor{black}{subsection}, we conduct the fine-grained prediction performance experiment on three datasets to evaluate the ability to eliminate the range effect. Compared methods includes HyperIQA\cite{HyperIQA}, MetaIQA\cite{MetaIQA}, and GraphIQA \cite{GraphIQA}. All the methods utilize \textcolor{black}{the} same division of the database to keep the consistency of test samples. Results from respective best models is evaluated in each category interval. For further examining the effectiveness of REQA, we also analyze the prediction bias and the number of outliers statistically.

\begin{table}[!t]
\setlength{\abovecaptionskip}{0.2cm}
\setlength{\belowcaptionskip}{0.2cm}
\centering
\caption{\textcolor{black}{Fine-grained} prediction performance on three datasets. All methods are conducted using identical training and testing \textcolor{black}{protocols}. The best results are highlighted in bold.}
\fontsize{8}{12}\selectfont
\begin{tabular}{c | ccccc }
\hline
SROCC  & Excellent  & Good & Fair & Poor & Bad \\
\hline
  MetaIQA\cite{MetaIQA}        &0.125 &0.699& 0.654 & 0.322 &0.426\\
  HyperIQA\cite{HyperIQA}    &0.167 &0.645 &0.560 &0.574 &0.519\\
  GraphIQA\cite{GraphIQA}  &0.133&0.688&0.624&0.517&\bfseries0.598\\
  REQA                     &\bfseries0.252&\bfseries0.751&\bfseries0.665&\bfseries0.722&0.571\\
\hline
\hline
PLCC  & Excellent  & Good & Fair & Poor & Bad \\
\hline
  MetaIQA\cite{MetaIQA}        &0.293&0.702&0.657& 0.363 & 0.180\\
  HyperIQA\cite{HyperIQA}    &0.289&0.640&0.567&0.585&0.507\\
  GraphIQA\cite{GraphIQA}  &0.311&0.688&0.629&0.563&0.570\\
\hline
  REQA&\bfseries0.435&\bfseries0.746&\bfseries0.660&\bfseries0.768&\bfseries0.851\\
\hline
\end{tabular}
\end{table}

% \begin{figure}[!t]
% \centering
% \includegraphics[width=1\columnwidth]{6}
% \caption{Statistics of prediction deviation \textcolor{black}{in terms of Absolute Category Rating (ACR) scales}.}
% \label{fig4}
% \end{figure}

% \begin{figure}[!t]
% \centering
% \includegraphics[width=1\columnwidth]{5}
% \caption{Statistics of prediction deviation \textcolor{black}{in terms of the prediction score}.}
% \label{fig4}
% \end{figure}

\subsubsection{Five quality ratings evaluation}
\textcolor{black}{As few methods focus on the fine-grained BIQA \cite{FGIQA}. There exists no benchmark to explicitly evaluate fine-grained ability. Here, we divide three datasets (i.e., CLIVE, KonIQ-10k, and BID) into five fine-grained quality ratings following \cite{BT500}: $Excellent$, $Good$, $Fair$, $Poor$, and $Bad$. }As is shown in Tab. \uppercase\expandafter{\romannumeral1}, REQA achieves outstanding performances in terms of all quality ratings. We can see that the range effect exists among almost all three methods: MetaIQA \cite{MetaIQA} performs well only on the $Good$ and $Fair$ quality ratings, which means if the input images are in the narrow range (i.e., $Excellent$, $Fair$, or $Bad$), the evaluation will degrade much; Also, HyperIQA \cite{HyperIQA} and GraphIQA\cite{GraphIQA} confront the same trouble that they both can not achieve ideal performance on the $Excellent$ and $Bad$ ratings. Compared with the three methods, REQA gets the best results in $Poor$, $Fair$, $Good$, and $Excellent$ ratings. Especially in $Poor$ rating, REQA achieves about $20\%$ improvement both on SROCC and PLCC in contrast to GraphIQA. In the $Excellent$ and $Bad$ ratings, four methods all fail to obtain ideal results. According to statistic analysis, there are a few samples in these two ranges, which greatly increases the prediction difficulty as even an outlier leads to the obvious disturbance on SROCC. In this case, REQA still keeps competitive. In the $Bad$ range, REQA achieves the best result on PLCC and the second-best result on SROCC. REQA also achieves the best in the $Excellent$ rating.

\begin{table}[t]\center
\begin{tabular}{c|c|c|c}
\hline
         & 0    & 1   & \textgreater{}=2 \\ \hline
REQA     & 1915 & 442 & 7                \\ 
GraphIQA \cite{GraphIQA} & 1796 & 559 & 19               \\ 
HyperIQA \cite{HyperIQA} & 1643 & 710 & 21               \\ 
MetaIQA  \cite{MetaIQA} & 1692 & 671 & 11               \\ \hline
\end{tabular}
\caption{Statistics of prediction deviation \textcolor{black}{in terms of Absolute Category Rating (ACR) scales}.}
\end{table}

\begin{table}[t]\center
\begin{tabular}{c|c|c|c|c|c}
\hline
         & {[}0,2.5{]} & {[}2.5,5{]} & {[}5,7.5{]} & \multicolumn{1}{l|}{{[}7.5, 10{]}} & \multicolumn{1}{l}{\textgreater{}=10} \\ \hline
REQA     & 918         & 390         & 431         & 237                                & 208                                   \\ 
GraphIQA \cite{GraphIQA} & 771         & 390         & 473         & 237                                & 265                                   \\ 
HyperIQA \cite{HyperIQA}& 655         & 372         & 440         & 296                                & 400                                   \\ 
MetaIQA \cite{MetaIQA} & 404         & 483         & 461         & 268                                & 213                                   \\ \hline
\end{tabular}
\caption{Statistics of prediction deviation \textcolor{black}{in terms of the prediction score}.}
\end{table}

\subsubsection{Outlier and deviation analysis}
To further verify the effectiveness of REQA, we make \textcolor{black}{a} statistic analysis between pMOS and the ground truth. As is shown in Fig. 5, we \textcolor{black}{depict} the experimental results on KonIQ-10K for direct visualization. On the whole, the linear fitting of the predicted results of REQA is more similar to the directly proportional function compared to MetaIQA\cite{MetaIQA} and HyperIQA\cite{HyperIQA}. Compared to GraphIQA\cite{GraphIQA}, the distribution of its predicted results is more concentrated. This demonstrates that the results of REQA possess the best linear correlation with the ground truth data on a wide range. Moreover, focusing on a particular range (\textcolor{black}{e.g.,} points within the green box), the results of REQA keep the same property. In addition, compared to the other methods, the predicted outliers (\textcolor{black}{e.g.,} points within \textcolor{black}{red} boxes) of REQA is more close to the fitting curve. This means that REQA can alleviate the prediction deviation from a wide range effectively.

For further analysis, the prediction deviation of different quality ratings is quantified as illustrated in Tab. \uppercase\expandafter{\romannumeral2}. Here, we illustrate the statistic quantity of samples in terms of the quality ratings difference of Absolute Category Rating (ACR)\cite{BT500} between pMOS and MOS. It is obvious that REQA is capable of limiting more images to its original rank scale, and meantime, it can reduce the wide range of prediction deviation as much as possible.

In addition, we also quantify the prediction deviation \textcolor{black}{in terms of prediction score [0, 100],} which is shown in Tab. \uppercase\expandafter{\romannumeral3}. Here, similar to Tab. \uppercase\expandafter{\romannumeral2}, we illustrate the number of the samples in terns of the prediction score differences of pMOS and MOS. Compared to the other methods, REQA achieves the best performance as there are $918$ samples in $[0, 2.5]$ and the most prediction biases concentrate in the first three ranges. This means that the predicted scores of REQA possess smaller fluctuation and furthermore demonstrates that the superiority of REQA for the fine-grained BIQA problem.

\subsection{\textcolor{black}{Coarse-grained} Prediction Performance Experiment}
In this \textcolor{black}{subsection}, we first conduct experiments on individual authentically distorted databases to verify the effectiveness of the proposed method \textcolor{black}{in terms of the traditional coarse-grained ability \cite{FGIQA}}\textcolor{black}{, and then we make statistically} significant test to validate the robustness of REQA. \textcolor{black}{Lastly, we explore the generalization ability of the proposed method.}
\subsubsection{Single database evaluations}
\begin{table}[t]
  \centering
  \caption{\textcolor{black}{Coarse-grained} prediction performance on authentically distorted databases CLIVE\cite{LIVEC}, KonIQ\cite{KonIQ} and BID\cite{BID}.The best two results are highlighted in bold.}
  \renewcommand\tabcolsep{4.0pt}
        \begin{tabular}{c | cc |cc| cc}
    \hline
     & \multicolumn{2}{c|}{CLIVE}  & \multicolumn{2}{c|}{BID}  & \multicolumn{2}{c}{KonIQ-10K} \\
    \hline
    IQA methods & SROCC  & PLCC   & SROCC  & PLCC  & SROCC  & PLCC    \\
    \hline
  BRISQUE\cite{BRISQUE}       &0.608&0.629&0.562&0.593&0.665&0.681 \\
  ILNIQE\cite{ILNIQE}               &0.432&0.508&0.516&0.554&0.507&0.523 \\
  HOSA\cite{HOSA}                 &0.640&0.678&0.721&0.736&0.671&0.694\\
  \hline
  BIECON\cite{BIECON}         &0.595&0.613&0.539&0.576&0.618&0.651\\
  WaDIQAM\cite{WaDIQaM}       &0.671&0.680 &0.725&0.742&0.797&0.805\\
  SFA\cite{SFA}               &0.812&0.833&0.826&0.840 &0.856&0.872\\
  PQR\cite{PQR}               &0.857&0.882&0.775&0.794&0.880 &0.884\\
  DBCNN\cite{DBCNN}           &0.851&0.869&0.845&0.859&0.875&0.884\\
  SGDNet\cite{SGD}                     &0.851&0.872&-&-&-&-\\
  MetaIQA\cite{MetaIQA}       &0.802&0.835&0.825&0.828&0.850& 0.887\\
  HyperIQA\cite{HyperIQA}     &\bfseries0.859&\bfseries0.882&0.869&\bfseries0.878&0.906&\bfseries0.917\\
  AIGQA\cite{AIGQA}           &0.751&0.761&-&-&-&-\\
  OLNet\cite{OLNet}           &0.849&0.858&-&-&0.877&0.882\\
  GraphIQA\cite{GraphIQA}           &0.845&0.862&\bfseries0.870&0.872&\bfseries0.911&0.915\\
 \hline
  REQA                        &\bfseries0.868&\bfseries0.880&\bfseries0.878&\bfseries0.889&\bfseries0.916&\bfseries0.920\\
    \hline
    \end{tabular}%
\end{table}%

% \begin{table}[!t]
% \setlength{\abovecaptionskip}{0.2cm}
% \setlength{\belowcaptionskip}{0.2cm}
%   \renewcommand\tabcolsep{4.0pt}
% \centering
% \caption{Statistic significance test results on CLIVE\cite{LIVEC}.}
% \fontsize{8}{12}\selectfont
% %\setlength{\tabcolsep}{4.5mm}
% \begin{tabular}{c | ccccc}
% \hline
% & AIGQA & DBCNN & HyperIQA & GraphIQA  & REQA \\
% \hline
%   AIGQA\cite{AIGQA}  &0 &-1&-1 & -1 &-1\\
%   DBCNN\cite{DBCNN}    &1 &0&-1&-1&-1 \\
%   HyperIQA\cite{HyperIQA}   &1 &1 &0 &-1 &-1\\
%   GraphIQA\cite{GraphIQA} &1&1&1&0&0\\
%   REQA     &1&1&1&0&0\\
% \hline
% \end{tabular}
% \end{table}

We compare the proposed REQA with 3 traditional methods and 11 DNN-based algorithms. The experimental results are exhibited in Tab. \uppercase\expandafter{\romannumeral4}, where the top two SROCC and PLCC are marked in bold. All the results of the traditional methods are implemented from the original codes. As for the DNN-based methods\cite{BIECON}\cite{WaDIQaM}\cite{SFA}\cite{PQR}\cite{DBCNN}\cite{SGD}\cite{MetaIQA}\cite{HyperIQA}\cite{AIGQA}\cite{OLNet}\cite{GraphIQA}, the results are taken from respective papers or reproduced by the source codes released by their authors.

Compared with the traditional methods, REQA possesses a notable advantage. Particularly in KonIQ, REQA achieves competitive performance, with \textcolor{black}{an} improvement of about $23\%$ on SROCC and $22\%$ on PLCC. It benefits from the powerful feature representation ability of DNN, which can obtain more diverse degradation information to perceive the quality of the image in the real world than the traditional methods can do.

When compared with DNN-based algorithms, REQA also achieves promising results on 3 databases. Specifically, REQA clearly outperforms all the methods on BID and KonIQ. As for CLIVE, REQA obtains the best result on SROCC, and on PLCC, it gets the second best results. Overall, the proposed method achieves \textcolor{black}{an} outstanding improvement of SROCC in experiments, which suggests that the design for alleviating the range effect enhances the sensitivity to the change of image quality, thereby optimizing the order of the predicted scores. 

\subsubsection{Generalization Ability Test}
In order to explore the generalization ability of the proposed model, we run \textcolor{black}{cross-database} tests on authentically distorted IQA databases compared with the other three methods DBCNN\cite{DBCNN}, HyperIQA\cite{HyperIQA}, and GraphIQA \cite{GraphIQA}. The experiment is conducted by training on one database and testing on the full of another database. As we can see in Tab. \uppercase\expandafter{\romannumeral5}, the proposed method achieves outstanding generalization ability as it obtains five best results and one top-two results. Especially compared to GraphIQA\cite{GraphIQA}, REQA achieves \textcolor{black}{a} $6.3\%$ improvement on SROCC when trained on CLIVE and tested on BID. The performance of HyperIQA\cite{HyperIQA} is better than ours in the setting of training on the BID and testing on KonIQ. The reason for this is that the self-adaptive strategy makes it effective to assess the image quality according to content information. In the case that the scene changes, it can quickly perceive this change. The advantage reminds us to perfect the proposed method by improving the adaptability in future work.

\begin{table}[!t]
  \setlength{\abovecaptionskip}{0.2cm}
  \setlength{\belowcaptionskip}{0.2cm}
  \renewcommand\tabcolsep{4.0pt}
\centering
\caption{SROCC results of the cross-database evaluations. The experiment is conducted by training on one database and testing on the full of another database.}
 \fontsize{8}{12}\selectfont
\begin{tabular}{c|c|c|c|c|c}
\hline
& Testing & DBCNN\cite{DBCNN}   &HyperIQA\cite{HyperIQA}  &GraphIQA\cite{GraphIQA}  & REQA \\ \hline
\multicolumn{1}{c|}{CLIVE} & BID     & 0.714 & 0.762 & 0.756    & \bfseries0.825   \\ \cline{2-6}
\multicolumn{1}{c|}{}      & KonIQ   & 0.757 & 0.754 & 0.762    & \bfseries0.772   \\ \hline
\multicolumn{1}{c|}{BID}   & CLIVE   & 0.680 & 0.725 & 0.747    & \bfseries0.770    \\ \cline{2-6}
\multicolumn{1}{c|}{}      & KonIQ   & 0.636 & \bfseries0.724 & 0.688    & 0.699   \\ \hline
\multicolumn{1}{c|}{KonIQ} & CLIVE   & 0.770 & 0.755 & 0.772    & \bfseries0.785   \\ \cline{2-6}
\multicolumn{1}{c|}{}      & BID     & 0.755 & 0.816 & 0.819    & \bfseries0.833   \\ \hline
\end{tabular}
\end{table}

\subsection{Ablation Study}
In this \textcolor{black}{subsection}, we make ablation experiments to verify the contribution of key constituent parts of REQA. We conduct \textcolor{black}{ablation experiments} on KonIQ \cite{KonIQ}. The training and testing protocols are the same as above.

\begin{table}[t]
  \setlength{\abovecaptionskip}{0.2cm}
  \setlength{\belowcaptionskip}{0.2cm}
  \renewcommand\tabcolsep{6.0pt}
\centering
\caption{Ablation Experiments about the performance of each module. \textcolor{black}{Here, we compare the performance of baseline (ResNet-50), TE, and FN based on the \textcolor{black}{fine-grained} prediction performance.}}
 \fontsize{8}{12}\selectfont
\begin{tabular}{l|ccccc}
\hline
SROOC            & Excellent & Good   & Fair  & Poor  & Bad   \\ \hline
w/o FN+TE   & 0.115      & 0.651   & 0.595  & 0.640  & 0.332   \\ \hline
w/o TE      & 0.238      & 0.726   & 0.642  & 0.703  & 0.560 \\ \hline
w/o FN      & 0.153      & 0.691   & 0.630  & 0.668  & 0.346 \\ 
w/ $t$=2    & 0.217      & 0.722   & 0.649  & 0.689  & 0.463 \\ 
w/ $t$=3    & 0.228      & 0.740   & 0.662  & 0.701  & 0.485 \\ 
w/ $t$=5    & 0.252      & 0.752   & 0.664  & 0.720  & 0.573 \\ \hline
REQA        & 0.252      & 0.751   & 0.665  & 0.722  & 0.571 \\ \hline\hline
PLCC        & Excellent & Good   & Fair  & Poor  & Bad   \\ \hline
w/o FN+TE   & 0.340      & 0.693   & 0.596  & 0.684  & 0.776   \\ \hline
w/o TE      & 0.425      & 0.732   & 0.643  & 0.754  & 0.838 \\ \hline
w/o FN      & 0.349      & 0.715   & 0.633  & 0.734  & 0.796 \\ 
w/ $t$=2    & 0.372      & 0.728   & 0.647  & 0.753  & 0.830 \\ 
w/ $t$=3    & 0.405      & 0.740   & 0.656  & 0.762  & 0.846 \\ 
w/ $t$=5    & 0.436      & 0.745   & 0.659  & 0.769  & 0.853 \\ \hline
REQA        & 0.435      & 0.746   & 0.660  & 0.768  & 0.851 \\ \hline
\end{tabular}
\end{table}

\begin{table}[t]
  \setlength{\abovecaptionskip}{0.2cm}
  \setlength{\belowcaptionskip}{0.2cm}
  \renewcommand\tabcolsep{6.0pt}
\centering
\caption{Ablation Experiments about the effectiveness of coarse-grained loss and fine-grained loss. The comparison is conducted based on the fine-grained} prediction performance.
 \fontsize{8}{12}\selectfont
\begin{tabular}{l|l|l|l|l|l}
\hline
SROCC                    & Excellent & Good   & Fair  & Poor  & Bad   \\ \hline
w/o $L^{Coarse}$   & 0.233     & 0.728     & 0.626   & 0.686  & 0.558 \\ \hline
w/o $L^{Fine}$     & 0.213     & 0.716     & 0.609   & 0.659  & 0.536 \\ \hline
REQA               & 0.252      & 0.751    & 0.665   & 0.722  & 0.571 \\ \hline\hline
PLCC               & Excellent  & Good     & Fair    & Poor   & Bad   \\ \hline
w/o $L^{Coarse}$   & 0.398      & 0.725    & 0.644   &0.743   & 0.827 \\ \hline
w/o $L^{Fine}$      & 0.365      & 0.704    & 0.629   &0.725   & 0.807 \\ \hline
REQA               & 0.435      & 0.746    & 0.660   &0.768   & 0.851 \\ \hline
\end{tabular}
\end{table}
We first examine the gains of three modules, the baseline network \textbf{ResNet-50}, \textbf{TE}, and \textbf{FN}. The results are listed in Tab. \uppercase\expandafter{\romannumeral6}. The modified model only removes the respective module, and the loss function is not changed. Specifically, in the model \textbf{w/o FN+TE}, the baseline network (\textbf{ResNet-50}) w/o FN+TE \cite{ResNet} is left to predict the image quality scores, which greatly degrades both SROCC and PLCC metrics. Compare to \textbf{w/o FN+TE}, model \textbf{w/o TE} slightly degrades to some extent due to lacking of global context information. As for the configuration of ablating FN (\textbf{w/o FN}), we can learn that feedback mechanism is conducive to different quality ratings, especially for $Excellent$ and $Bad$. Moreover, to further analyze the effectiveness of the feedback mechanism in FN module, we change the number of time steps (i.e., \textbf{w/ t=2;3;5}). We can see that with the increasing of number of time steps, the performance gains in terms of $Excellent$ and $Bad$ quality ratings are more than other quality ratings, which validates that the feedback mechanism can gradually refine the fine-grained prediction results. Besides, \textbf{w/ t=4} reaches the optimal point between performance and computation efficiency. Our proposed model (REQA) achieves the best results, which illustrates that modules interact with each other to get a positive gain.

Then we analyze the effectiveness of $L^{Coarse}$ and ${L^{Fine}}$, which is shown in Tab. \uppercase\expandafter{\romannumeral7}. We replace the ${L^{Coarse}}$ (w/o $L^{Coarse}$) and $L^{Fine}$ (w/o $L^{Fine}$) with $L_1$ loss, respectively. According to the fine-grained performance comparisons, we can conclude that coarse-to-fine loss functions, especially for $L^{Fine}$, benefit from different quality ratings.

Overall, we can make the subsequent conclusions. First of all, different modules proposed in this paper can improve the performance of REQA. \textbf{TE} enlarges the receptive field to acquire the context features. \textbf{FN} can obtain a more quality-aware perception of multi-scale features while completing the task iterations. $\bm{L^{Fine}}$ makes REQA have a prediction accuracy and with the addition of $\bm{L^{Coarse}}$, fine-grained prediction performance gets promoted. All of these are responsible for the outstanding experimental results of REQA.

\begin{figure}[t]
\centering
\includegraphics[width=1\columnwidth]{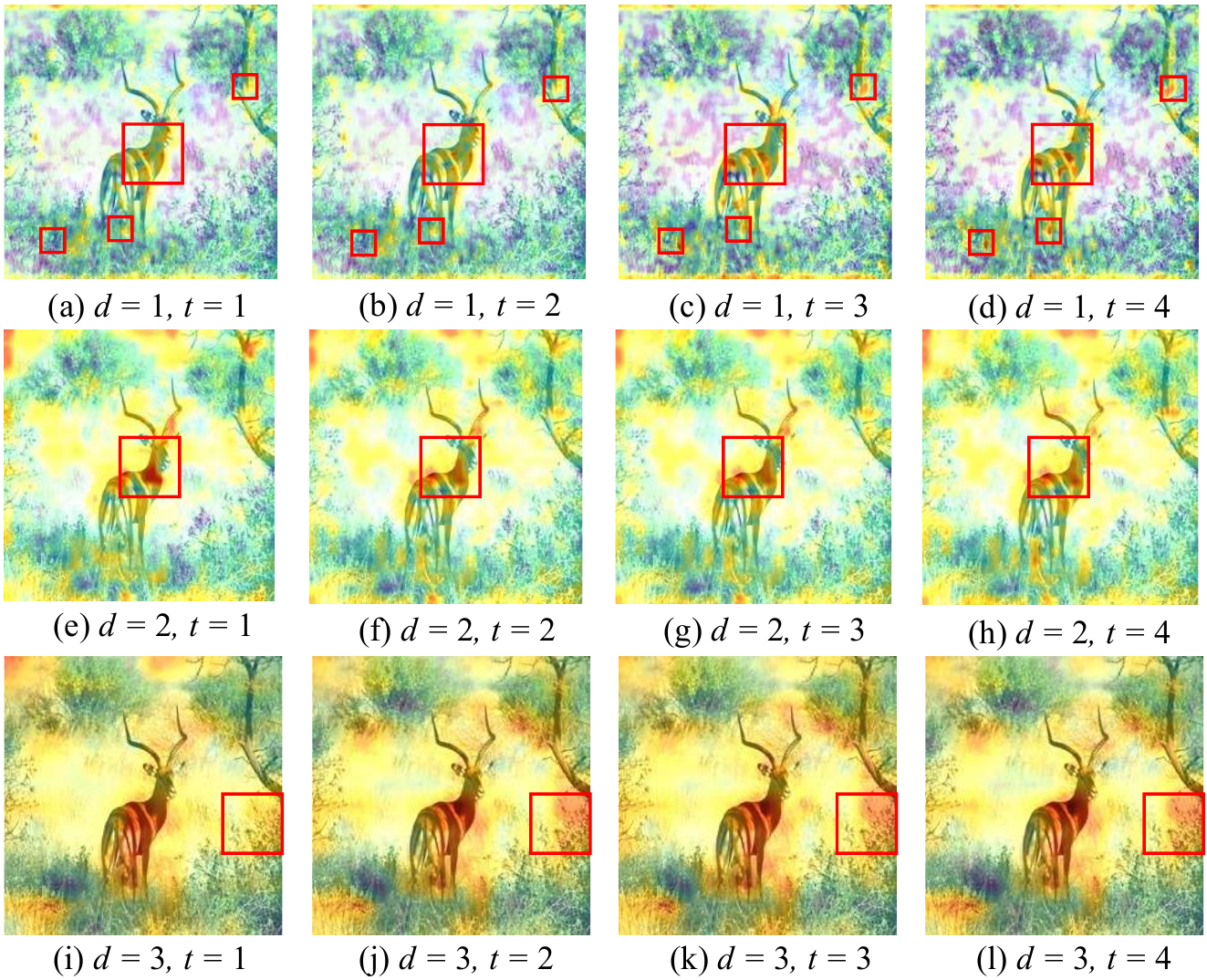}
\caption{Visualization of the fine-grained distortion information in the form of the heat map.}
\end{figure}
\subsection{Visualization of Fine-grained Distortion Features}
In this subsection, we visualize the fine-grained distortion information in the form of the heat map\cite{GC} to further examine the performance of the feedback hierarchy.
We show each heat map corresponding to the output of $d'th$ feedback block in $t'th$ time step in Fig. 6. Here, the displayed images are from KonIQ\cite{KonIQ}. As we can see from the longitudinal comparison, different feedback blocks possess different receptive fields. For example, by comparing Fig. 6 (a), (e), and (i), it can be concluded that the third feedback block perceives larger areas where the textures are much clearer.
This is attributed to different scales of inputs of three feedback blocks. Furthermore, by horizontal comparison, it can be concluded that feedback blocks can obtain more fine-grained feature maps, which is consistent with our standpoint. For example, through making comparisons of Fig. 6 (a)-(d), we can see that the heat map in $t=4$ has more heat sensitive areas (e.g., points within green boxes). In addition, we obtain the pMOS of the image in each time step as $79.01$ in $t=1$, $82.02$ in $t = 2$, $83.16$ in $t = 3$, and $82.13$ in $t = 4$. The MOS of it is $84.91$. This means that with the feedback going on, REQA can make a more accurate prediction. Overall, we can conclude that the feedback hierarchy can achieve better and better features as iteration goes on. These features are quality-aware enough to ensure REQA obtain more fine-grained prediction result.

\section{Conclusion}
 In this paper, to our best knowledge, we take the first attempt to develop a fine-grained blind image quality assessment
method to alleviate the range effect.
Concretely, we first propose the coarse-to-fine method with a strong fine-grained prediction ability. Benefiting from Feedback Network (FN) and  Transformer Encoder (TE), the proposed method can perceive the multi-scale distortion information and global context information, which makes the model quality-aware for fine-grained distortions. Furthermore, by integrating coarse-grained metric and fine-grained losses into the feedback hierarchy to process these features, the proposed method achieves outstanding coarse-grained and fine-grained prediction performance, as is demonstrated by a series of experimental results. In future work, we will try to develop a metric to quantitatively evaluate the range effect via statistical methods, which can also be applied to evaluate the fine-grained prediction ability of existing IQA methods.

{
\bibliographystyle{IEEEtran}
\bibliography{egbib.bib}

\begin{thebibliography}{10}
\providecommand{\url}[1]{#1}
\csname url@samestyle\endcsname
\providecommand{\newblock}{\relax}
\providecommand{\bibinfo}[2]{#2}
\providecommand{\BIBentrySTDinterwordspacing}{\spaceskip=0pt\relax}
\providecommand{\BIBentryALTinterwordstretchfactor}{4}
\providecommand{\BIBentryALTinterwordspacing}{\spaceskip=\fontdimen2\font plus
\BIBentryALTinterwordstretchfactor\fontdimen3\font minus
  \fontdimen4\font\relax}
\providecommand{\BIBforeignlanguage}[2]{{%
\expandafter\ifx\csname l@#1\endcsname\relax
\typeout{** WARNING: IEEEtran.bst: No hyphenation pattern has been}%
\typeout{** loaded for the language `#1'. Using the pattern for}%
\typeout{** the default language instead.}%
\else
\language=\csname l@#1\endcsname
\fi
#2}}
\providecommand{\BIBdecl}{\relax}
\BIBdecl

\bibitem{image_restoration}
Z.~Pan, F.~Yuan, J.~Lei, Y.~Fang, X.~Shao, and S.~Kwong, ``Vcrnet: Visual
  compensation restoration network for no-reference image quality assessment,''
  \emph{IEEE TIP}, vol.~31, pp. 1613--1627, 2022.

\bibitem{image_compression}
J.~Chang, Z.~Zhao, C.~Jia, S.~Wang, L.~Yang, Q.~Mao, J.~Zhang, and S.~Ma,
  ``Conceptual compression via deep structure and texture synthesis,''
  \emph{IEEE TIP}, vol.~31, pp. 2809--2823, 2022.

\bibitem{IEEETETCI3}
R.~Tu, G.~Jiang, M.~Yu, T.~Luo, Z.~Peng, and F.~Chen, ``V-pcc projection based
  blind point cloud quality assessment for compression distortion,'' \emph{IEEE
  Transactions on Emerging Topics in Computational Intelligence}, vol.~7,
  no.~2, pp. 462--473, 2023.

\bibitem{SSIM}
Z.~Wang, A.~C. Bovik, H.~R. Sheikh, E.~P. Simoncelli \emph{et~al.}, ``Image
  quality assessment: from error visibility to structural similarity,''
  \emph{IEEE TIP}, vol.~13, no.~4, pp. 600--612, 2004.

\bibitem{VIF}
H.~R. Sheikh and A.~C. Bovik, ``Image information and visual quality,''
  \emph{IEEE TIP}, vol.~15, no.~2, pp. 430--444, 2006.

\bibitem{FSIM}
L.~Zhang, L.~Zhang, X.~Mou, and D.~Zhang, ``Fsim: A feature similarity index
  for image quality assessment,'' \emph{IEEE TIP}, vol.~20, no.~8, pp.
  2378--2386, 2011.

\bibitem{RR1}
L.~Ma, S.~Li, F.~Zhang, and K.~N. Ngan, ``Reduced-reference image quality
  assessment using reorganized dct-based image representation,'' \emph{IEEE
  TMM}, vol.~13, no.~4, pp. 824--829, 2011.

\bibitem{RR2}
G.~Zhai, X.~Wu, X.~Yang, W.~Lin, and W.~Zhang, ``A psychovisual quality metric
  in free-energy principle,'' \emph{IEEE TIP}, vol.~21, no.~1, pp. 41--52,
  2011.

\bibitem{RR3}
A.~Rehman and Z.~Wang, ``Reduced-reference image quality assessment by
  structural similarity estimation,'' \emph{IEEE TIP}, vol.~21, no.~8, pp.
  3378--3389, 2012.

\bibitem{HyperIQA}
S.~Su, Q.~Yan, Y.~Zhu, C.~Zhang, X.~Ge, J.~Sun, and Y.~Zhang, ``Blindly assess
  image quality in the wild guided by a self-adaptive hyper network,'' in
  \emph{CVPR}, 2020, pp. 3667--3676.

\bibitem{LIVEC}
D.~Ghadiyaram and A.~C. Bovik, ``Massive online crowdsourced study of
  subjective and objective picture quality,'' \emph{IEEE TIP}, vol.~25, no.~1,
  pp. 372--387, 2016.

\bibitem{CORNIA}
P.~Ye, J.~Kumar, L.~Kang, and D.~Doermann, ``Unsupervised feature learning
  framework for no-reference image quality assessment,'' in \emph{CVPR}, 2012,
  pp. 1098--1105.

\bibitem{DeepSim}
F.~Gao, Y.~Wang, P.~Li, M.~Tan, J.~Yu, and Y.~Zhu, ``Deepsim: Deep similarity
  for image quality assessment,'' \emph{Neurocomputing}, vol. 257, pp.
  104--114, 2017.

\bibitem{WaDIQaM}
S.~Bosse, D.~Maniry, K.-R. M{\"u}ller, T.~Wiegand, and W.~Samek, ``Deep neural
  networks for no-reference and full-reference image quality assessment,''
  \emph{IEEE TIP}, vol.~27, no.~1, pp. 206--219, 2017.

\bibitem{MEON}
K.~Ma, W.~Liu, K.~Zhang, Z.~Duanmu, Z.~Wang, and W.~Zuo, ``End-to-end blind
  image quality assessment using deep neural networks,'' \emph{IEEE TIP},
  vol.~27, no.~3, pp. 1202--1213, 2018.

\bibitem{dipIQ}
K.~Ma, W.~Liu, T.~Liu, Z.~Wang, and D.~Tao, ``dipiq: Blind image quality
  assessment by learning-to-rank discriminable image pairs,'' \emph{IEEE TIP},
  vol.~26, no.~8, pp. 3951--3964, 2017.

\bibitem{RRLRIQA}
J.~Gu, G.~Meng, C.~Da, S.~Xiang, and C.~Pan, ``No-reference image quality
  assessment with reinforcement recursive list-wise ranking,'' in \emph{AAAI},
  vol.~33, no.~01, 2019, pp. 8336--8343.

\bibitem{DeepQA}
J.~Kim, A.-D. Nguyen, and S.~Lee, ``Deep cnn-based blind image quality
  predictor,'' \emph{IEEE TNNLS}, vol.~30, no.~1, pp. 11--24, 2018.

\bibitem{BPSQM}
D.~Pan, P.~Shi, M.~Hou, Z.~Ying, S.~Fu, and Y.~Zhang, ``Blind predicting
  similar quality map for image quality assessment,'' in \emph{CVPR}, 2018, pp.
  6373--6382.

\bibitem{HVSIQA}
S.~Seo, S.~Ki, and M.~Kim, ``Deep hvs-iqa net: Human visual system inspired
  deep image quality assessment networks,'' \emph{arXiv preprint
  arXiv:1902.05316}, 2019.

\bibitem{HIQA}
K.-Y. Lin and G.~Wang, ``Hallucinated-iqa: No-reference image quality
  assessment via adversarial learning,'' in \emph{CVPR}, 2018, pp. 732--741.

\bibitem{RAN4IQA}
H.~Ren, D.~Chen, and Y.~Wang, ``Ran4iqa: Restorative adversarial nets for
  no-reference image quality assessment,'' in \emph{AAAI}, 2018.

\bibitem{SIQA}
L.~Shen, R.~Fang, Y.~Yao, X.~Geng, and D.~Wu, ``No-reference stereoscopic image
  quality assessment based on image distortion and stereo perceptual
  information,'' \emph{IEEE Transactions on Emerging Topics in Computational
  Intelligence}, vol.~3, no.~1, pp. 59--72, 2019.

\bibitem{RankIQA}
X.~Liu, J.~van~de Weijer, and A.~D. Bagdanov, ``Rankiqa: Learning from rankings
  for no-reference image quality assessment,'' in \emph{ICCV}, 2017, pp.
  1040--1049.

\bibitem{LIVEI}
H.~R. Sheikh, M.~F. Sabir, and A.~C. Bovik, ``A statistical evaluation of
  recent full reference image quality assessment algorithms,'' \emph{IEEE TIP},
  vol.~15, no.~11, pp. 3440--3451, 2006.

\bibitem{LIVEII}
D.~Jayaraman, A.~Mittal, A.~K. Moorthy, and A.~C. Bovik, ``Objective quality
  assessment of multiply distorted images,'' in \emph{2012 Conference record of
  the forty sixth asilomar conference on signals, systems and computers
  (ASILOMAR)}, 2012, pp. 1693--1697.

\bibitem{TID2013}
N.~Ponomarenko, L.~Jin, O.~Ieremeiev, V.~Lukin, K.~Egiazarian, J.~Astola,
  B.~Vozel, K.~Chehdi, M.~Carli, F.~Battisti \emph{et~al.}, ``Image database
  tid2013: Peculiarities, results and perspectives,'' \emph{Signal Processing:
  Image Communication}, vol.~30, pp. 57--77, 2015.

\bibitem{CSIQ}
E.~C. Larson and D.~M. Chandler, ``Most apparent distortion: full-reference
  image quality assessment and the role of strategy,'' \emph{Journal of
  Electronic Imaging}, vol.~19, no.~1, p. 011006, 2010.

\bibitem{KADID}
H.~Lin, V.~Hosu, and D.~Saupe, ``Kadid-10k: A large-scale artificially
  distorted iqa database,'' in \emph{QoMEX}, 2019, pp. 1--3.

\bibitem{DBCNN}
W.~Zhang, K.~Ma, J.~Yan, D.~Deng, and Z.~Wang, ``Blind image quality assessment
  using a deep bilinear convolutional neural network,'' \emph{IEEE TCSVT},
  vol.~30, no.~1, pp. 36--47, 2018.

\bibitem{BLINDER}
F.~Gao, J.~Yu, S.~Zhu, Q.~Huang, and Q.~Tian, ``Blind image quality prediction
  by exploiting multi-level deep representations,'' \emph{Pattern Recognition},
  vol.~81, pp. 432--442, 2018.

\bibitem{MetaIQA}
H.~Zhu, L.~Li, J.~Wu, W.~Dong, and G.~Shi, ``Metaiqa: Deep meta-learning for
  no-reference image quality assessment,'' in \emph{CVPR}, 2020, pp.
  14\,143--14\,152.

\bibitem{SFA}
D.~Li, T.~Jiang, W.~Lin, and M.~Jiang, ``Which has better visual quality: The
  clear blue sky or a blurry animal?'' \emph{IEEE TMM}, vol.~21, no.~5, pp.
  1221--1234, 2018.

\bibitem{IEEETETCI1}
L.~Shen, R.~Fang, Y.~Yao, X.~Geng, and D.~Wu, ``No-reference stereoscopic image
  quality assessment based on image distortion and stereo perceptual
  information,'' \emph{IEEE Transactions on Emerging Topics in Computational
  Intelligence}, vol.~3, no.~1, pp. 59--72, 2019.

\bibitem{NARCNN}
Y.~Liang, J.~Wang, X.~Wan, Y.~Gong, and N.~Zheng, ``Image quality assessment
  using similar scene as reference,'' in \emph{ECCV}, 2016, pp. 3--18.

\bibitem{DeepRN}
D.~Varga, D.~Saupe, and T.~Szir{\'a}nyi, ``Deeprn: A content preserving deep
  architecture for blind image quality assessment,'' in \emph{ICME}, 2018, pp.
  1--6.

\bibitem{PQR}
H.~Zeng, L.~Zhang, and A.~C. Bovik, ``Blind image quality assessment with a
  probabilistic quality representation,'' in \emph{ICIP}.\hskip 1em plus 0.5em
  minus 0.4em\relax IEEE, 2018, pp. 609--613.

\bibitem{GraphIQA}
S.~Sun, T.~Yu, J.~Xu, W.~Zhou, and Z.~Chen, ``Graphiqa: Learning distortion
  graph representations for blind image quality assessment,'' \emph{IEEE TMM},
  pp. 1--1, 2022.

\bibitem{IEEETETCI2}
Y.~Cui, G.~Jiang, M.~Yu, and Y.~Song, ``Local visual and global deep features
  based blind stitched panoramic image quality evaluation using ensemble
  learning,'' \emph{IEEE Transactions on Emerging Topics in Computational
  Intelligence}, vol.~6, no.~5, pp. 1222--1236, 2022.

\bibitem{TMM}
Z.~Wang, Q.~Jiang, S.~Zhao, W.~Feng, and W.~Lin, ``Deep blind image quality
  assessment powered by online hard example mining,'' \emph{IEEE TMM}, pp.
  1--11, 2023.

\bibitem{KonIQ}
V.~Hosu, H.~Lin, T.~Sziranyi, and D.~Saupe, ``Koniq-10k: An ecologically valid
  database for deep learning of blind image quality assessment,'' \emph{IEEE
  TIP}, vol.~29, pp. 4041--4056, 2020.

\bibitem{BID}
A.~Ciancio, E.~A. da~Silva, A.~Said, R.~Samadani, P.~Obrador \emph{et~al.},
  ``No-reference blur assessment of digital pictures based on multifeature
  classifiers,'' \emph{IEEE TIP}, vol.~20, no.~1, pp. 64--75, 2010.

\bibitem{RE}
M.~A. Saad, P.~Le~Callet, and P.~Corriveau, ``Blind image quality assessment:
  Unanswered questions and future directions in the light of consumers needs,''
  \emph{VQEG eLetter}, vol.~1, no.~2, pp. 62--66, 2014.

\bibitem{range_effect1}
L.~Krasula, P.~Le~Callet, K.~Fliegel, and M.~Klíma, ``Quality assessment of
  sharpened images: Challenges, methodology, and objective metrics,''
  \emph{IEEE TIP}, vol.~26, no.~3, pp. 1496--1508, 2017.

\bibitem{REM}
L.~Krasula, K.~Fliegel, P.~Le~Callet, and M.~Kl{\'\i}ma, ``On the accuracy of
  objective image and video quality models: New methodology for performance
  evaluation,'' in \emph{QoMEX}, 2016, pp. 1--6.

\bibitem{PaQ}
Z.~Ying, H.~Niu, P.~Gupta, D.~Mahajan, D.~Ghadiyaram, and A.~Bovik, ``From
  patches to pictures (paq-2-piq): Mapping the perceptual space of picture
  quality,'' in \emph{CVPR}, 2020, pp. 3575--3585.

\bibitem{FGIQA}
X.~Zhang, W.~Lin, and Q.~Huang, ``Fine-grained image quality assessment: A
  revisit and further thinking,'' \emph{IEEE TCSVT}, vol.~32, no.~5, pp.
  2746--2759, 2022.

\bibitem{PieAPP}
E.~Prashnani, H.~Cai, Y.~Mostofi, and P.~Sen, ``Pieapp: Perceptual image-error
  assessment through pairwise preference,'' in \emph{CVPR}, 2018, pp.
  1808--1817.

\bibitem{ML}
A.~Bellet, A.~Habrard, and M.~Sebban, ``Metric learning,'' \emph{Synthesis
  Lectures on Artificial Intelligence and Machine Learning}, vol.~9, no.~1, pp.
  1--151, 2015.

\bibitem{PL}
R.~L. Goldstone, ``Perceptual learning,'' \emph{Annual review of psychology},
  vol.~49, no.~1, pp. 585--612, 1998.

\bibitem{Feedback1}
S.~Hochstein and M.~Ahissar, ``View from the top: Hierarchies and reverse
  hierarchies in the visual system,'' \emph{Neuron}, vol.~36, no.~5, pp.
  791--804, 2002.

\bibitem{FeedBack2}
A.~R. Zamir, T.-L. Wu, L.~Sun, W.~B. Shen, B.~E. Shi, J.~Malik, and
  S.~Savarese, ``Feedback networks,'' in \emph{CVPR}, 2017, pp. 1308--1317.

\bibitem{CL}
Y.~Bengio, J.~Louradour, R.~Collobert, and J.~Weston, ``Curriculum learning,''
  in \emph{ICML}, 2009, pp. 41--48.

\bibitem{DIIVINE}
A.~K. Moorthy and A.~C. Bovik, ``Blind image quality assessment: From natural
  scene statistics to perceptual quality,'' \emph{IEEE TIP}, vol.~20, no.~12,
  pp. 3350--3364, 2011.

\bibitem{BLIINDSII}
M.~A. Saad, A.~C. Bovik, and C.~Charrier, ``Blind image quality assessment: A
  natural scene statistics approach in the dct domain,'' \emph{IEEE TIP},
  vol.~21, no.~8, pp. 3339--3352, 2012.

\bibitem{BRISQUE}
A.~Mittal, A.~K. Moorthy, and A.~C. Bovik, ``No-reference image quality
  assessment in the spatial domain,'' \emph{IEEE TIP}, vol.~21, no.~12, pp.
  4695--4708, 2012.

\bibitem{NRSL}
Q.~Li, W.~Lin, J.~Xu, and Y.~Fang, ``Blind image quality assessment using
  statistical structural and luminance features,'' \emph{IEEE TMM}, vol.~18,
  no.~12, pp. 2457--2469, 2016.

\bibitem{RISE}
L.~Li, W.~Xia, W.~Lin, Y.~Fang, and S.~Wang, ``No-reference and robust image
  sharpness evaluation based on multiscale spatial and spectral features,''
  \emph{IEEE TMM}, vol.~19, no.~5, pp. 1030--1040, 2016.

\bibitem{CBIQ}
P.~Ye and D.~Doermann, ``No-reference image quality assessment using visual
  codebooks,'' \emph{IEEE TIP}, vol.~21, no.~7, pp. 3129--3138, 2012.

\bibitem{JND}
S.-H. Bae and M.~Kim, ``A dct-based total jnd profile for spatiotemporal and
  foveated masking effects,'' \emph{IEEE TCSVT}, vol.~27, no.~6, pp.
  1196--1207, 2016.

\bibitem{GAN1}
X.~Shi, M.~Zhang, S.~Xia, R.~Zhang, and J.~Feng, ``Local feature enhanced
  adversarial network for the blind image quality assessment,'' in
  \emph{ICASSP}, 2023, pp. 1--5.

\bibitem{AIGQA}
J.~Ma, J.~Wu, L.~Li, W.~Dong, X.~Xie, G.~Shi, and W.~Lin, ``Blind image quality
  assessment with active inference,'' \emph{IEEE TIP}, vol.~30, pp. 3650--3663,
  2021.

\bibitem{ICME}
P.~Zhang, X.~Shao, and Z.~Li, ``Cycleiqa: Blind image quality assessment via
  cycle-consistent adversarial networks,'' in \emph{ICME}, 2022, pp. 1--6.

\bibitem{MAML}
C.~Finn, P.~Abbeel, and S.~Levine, ``Model-agnostic meta-learning for fast
  adaptation of deep networks,'' in \emph{ICML}, 2017, pp. 1126--1135.

\bibitem{TE}
A.~Vaswani, N.~Shazeer, N.~Parmar, J.~Uszkoreit, L.~Jones, A.~N. Gomez,
  {\L}.~Kaiser, and I.~Polosukhin, ``Attention is all you need,'' in
  \emph{NeurIPS}, 2017, pp. 5998--6008.

\bibitem{ResNet}
K.~He, X.~Zhang, S.~Ren, and J.~Sun, ``Deep residual learning for image
  recognition,'' in \emph{CVPR}, 2016, pp. 770--778.

\bibitem{SGD}
S.~Yang, Q.~Jiang, W.~Lin, and Y.~Wang, ``Sgdnet: An end-to-end saliency-guided
  deep neural network for no-reference image quality assessment,'' in \emph{ACM
  MM}, 2019, pp. 1383--1391.

\bibitem{ConvLSTM}
S.~Xingjian, Z.~Chen, H.~Wang, D.-Y. Yeung, W.-K. Wong, and W.-c. Woo,
  ``Convolutional lstm network: A machine learning approach for precipitation
  nowcasting,'' in \emph{NeurIPS}, 2015, pp. 802--810.

\bibitem{BERT}
J.~Devlin, M.-W. Chang, K.~Lee, and K.~Toutanova, ``Bert: Pre-training of deep
  bidirectional transformers for language understanding,'' \emph{arXiv preprint
  arXiv:1810.04805}, 2018.

\bibitem{VIT}
A.~Dosovitskiy, L.~Beyer, A.~Kolesnikov, D.~Weissenborn, X.~Zhai,
  T.~Unterthiner, M.~Dehghani, M.~Minderer, G.~Heigold, S.~Gelly \emph{et~al.},
  ``An image is worth 16x16 words: Transformers for image recognition at
  scale,'' \emph{arXiv preprint arXiv:2010.11929}, 2020.

\bibitem{tcsvt5}
B.~Li, W.~Zhang, M.~Tian, G.~Zhai, and X.~Wang, ``Blindly assess quality of
  in-the-wild videos via quality-aware pre-training and motion perception,''
  \emph{IEEE TCSVT}, vol.~32, no.~9, pp. 5944--5958, 2022.

\bibitem{BT500}
R.~I.-R. BT, ``Methodology for the subjective assessment of the quality of
  television pictures,'' \emph{International Telecommunication Union}, 2002.

\bibitem{ILNIQE}
L.~Zhang, L.~Zhang, and A.~C. Bovik, ``A feature-enriched completely blind
  image quality evaluator,'' \emph{IEEE TIP}, vol.~24, no.~8, pp. 2579--2591,
  2015.

\bibitem{HOSA}
J.~Xu, P.~Ye, Q.~Li, H.~Du, Y.~Liu, and D.~Doermann, ``Blind image quality
  assessment based on high order statistics aggregation,'' \emph{IEEE TIP},
  vol.~25, no.~9, pp. 4444--4457, 2016.

\bibitem{BIECON}
J.~Kim and S.~Lee, ``Fully deep blind image quality predictor,'' \emph{IEEE
  JSTSP}, vol.~11, no.~1, pp. 206--220, 2016.

\bibitem{OLNet}
X.~Yao, Q.~Cao, X.~Feng, G.~Cheng, and J.~Han, ``Learning to assess image
  quality like an observer,'' \emph{IEEE TNNLS}, pp. 1--13, 2022.

\bibitem{GC}
R.~R. Selvaraju, M.~Cogswell, A.~Das, R.~Vedantam, D.~Parikh, and D.~Batra,
  ``Grad-cam: Visual explanations from deep networks via gradient-based
  localization,'' in \emph{ICCV}, 2017, pp. 618--626.

\end{thebibliography}
}

\vspace{-1cm}
% \begin{IEEEbiography}[{\includegraphics[width=1in,height=1.25in,clip,keepaspectratio]{Biography/BinghengLi.jpg}}]{Bingheng Li}
% received the B.S. degree from the School of Telecommunication and Information Engineering, Xi'an University of Posts and Telecommunications, Xi¡¯an,China, in 2019, where he is currently pursuing the master¡¯s degree with the School of Electronic Engineering, Xidian University, Xi'an, China. His current research interests include machine learning, visual information processing and image quality assessment.
% \end{IEEEbiography}

% \enlargethispage{-11.5cm}

\end{document}